\documentclass[twocolumn,aps,superscriptaddress,prl]{revtex4-1}

\usepackage[caption=false]{subfig}
\usepackage{dcolumn}
\usepackage{amsmath}
\usepackage{bm}
\usepackage{color}
\usepackage[english]{babel}
\usepackage{calrsfs}
\DeclareMathAlphabet{\pazocal}{OMS}{zplm}{m}{n}

\usepackage{graphicx,psfrag} 
\usepackage{natbib}
\usepackage{epstopdf}
\usepackage{appendix}
\usepackage{changes}

\definecolor{mygrey}{gray}{0.35}
\definecolor{myblue}{rgb}{0.2,0.2,0.8}
\definecolor{myzard}{cmyk}{0,0,0.05,0}
\definecolor{mywhite}{rgb}{1,1,1}
\definecolor{myred}{rgb}{1,0.,0.3}

\usepackage[colorlinks=true,citecolor=myblue,linkcolor=myred]{hyperref}

\def\be{\begin{equation}}
\def\ee{\end{equation}}
\def\ba{\begin{align}}
\def\enda{\end{align}}
\def\bi{\begin{itemize}}
\def\ei{\end{itemize}}

\def\beq{\begin{equation}}
\def\beq{\begin{equation}}
\def\eeq{\end{equation}}

 \newcommand{\ket}[1]{|#1\rangle}
 \newcommand{\bra}[1]{\langle #1|}

\graphicspath{{./Figures/}}

\begin{document}
\title[Short Title]{Achieving the ultimate precision limit in quantum NMR spectroscopy}
%\title[Short Title]{Achieving the SQL in quantum NMR spectroscopy without single nucleus addressing}
\author{D. Cohen}
\email{email: daniel.cohen7@mail.huji.ac.il}
\author{T. Gefen}
\author{L. Ortiz}
\author{A. Retzker}
\affiliation{Racah Institute of Physics, The Hebrew University of Jerusalem, Jerusalem 
	91904, Givat Ram, Israel}

	\begin{abstract}
The ultimate precision limit in estimating the Larmor frequency of $N$ unentangled rotating spins is well established, and is highly important for magnetometers, gyroscopes and many other sensors.
However this limit assumes perfect, single addressing, measurements of the spins. 
This requirement is not practical in NMR spectroscopy, as well as other physical systems, where a weakly interacting external probe is used as a measurement device.
Here we show that in the framework of quantum nano-NMR spectroscopy, in which these limitations are inherent, the ultimate precision limit is still achievable using control and a finely tuned measurement.     
	%maybe add weak interaction%      
	\end{abstract}
	\maketitle
	\section{INTRODUCTION}
	Nuclear Magnetic Resonance (NMR) spectroscopy is a revolutionary field that has had a considerable impact on science and medicine alike. It enables chemists to deduce molecular proprieties, and is best known for magnetic resonance imaging.
	The main idea behind the technology is that applying a large external magnetic field creates a macroscopic magnetization of the test sample. When a resonance pulse is implemented, the magnetization vector is transferred into the transverse plane, where it precesses around the external magnetic field axis. The oscillatory magnetic field of the sample can be sensed classically, and its spectral decomposition carries the target information\cite{Abragam1961,Slichter,NMR1,NMR2,NMR3}. 
	
\begin{figure}[t]
	\begin{center}
		\includegraphics[width=0.48\textwidth]{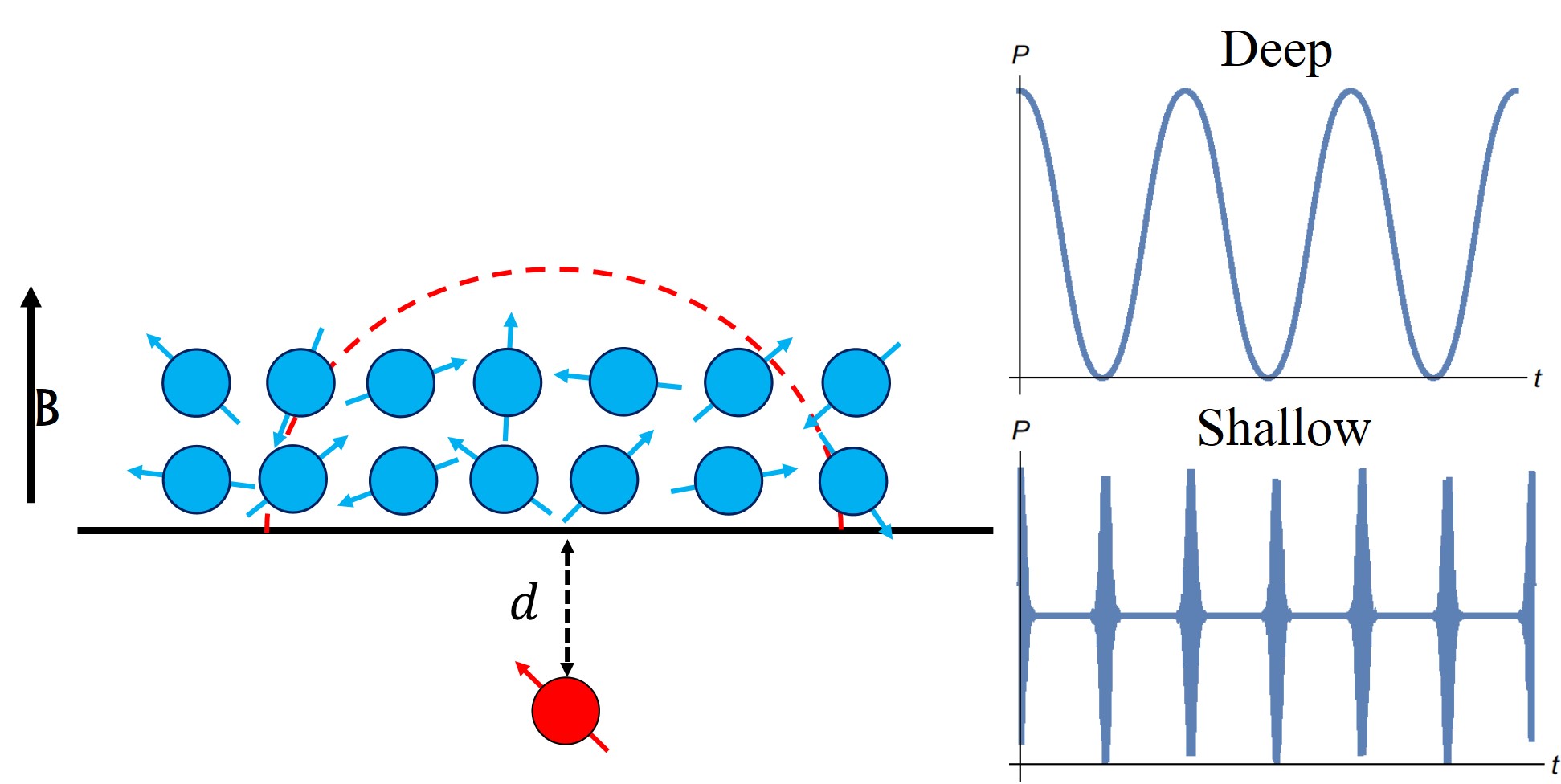}
	\end{center}
		\caption{ The NV center (red) is situated at a depth $d$ below the diamond surface. The nuclear spin ensemble (blue), which is located on top of the surface, is partially polarized due to the external magnetic field $B$. The local magnetic field at the NV's position is only affected by the nuclear spins within a hemisphere of radius $~d$ centered above the NV's position (dashed red line), since the dipolar interaction creates an effective cut off. The sensor is initialized and then measured after a given time, in order to reproduce the probability distribution which is analogous to the classic NMR signal (top right). However, as the sensor is brought into close proximity with the sample, strong back-action causes the signal to change (bottom right). This change dramatically affects the precision of the frequency estimation.}
		\label{System}
	\end{figure}

	Nano-scale NMR is a recent extension of this idea, that focuses on the measurement of the NMR spectra of minute samples. The main challenge stems from the fact that the net polarization of these samples cannot be detected using macroscopic devices. While various directions are being pursued to confront this challenge, one of the most promising is the Nitrogen-Vacancy (NV) based spectrometer which relies on a quantum emitter as a nano scale detector. The NV center is a natural candidate, since in the past decade it has been shown to be an excellent magnetometer on the nano-scale \cite{Review_NV,NV_Mag1}, and many successful experiments have been carried out in platform similar to fig. \ref{System} \cite{Staudacher,NanoNMRHarward,NanoNMRIBM,NanoNMRMeriles,NanoNMRQdyne,NanoNMRstutgart,Glenn}. To date, most ensemble experiments in the field have been conducted with bulk NVs, since surface effects and finite polarization limit the sensitivity  attained by the shallow ones. These experiments were able to measure impressively narrow spectra\cite{Glenn}. However, since the effective sample volume scales with the NV's depth, the use of shallow NVs is inevitable if one aims to interrogate ultra-small samples.           
	
	Whether the sensing is quantum or classic, the acquired signal is
	\beq\label{P_classic}
	S \sim NA\cos\left(\omega_N t\right),
	\eeq
	where $N$ is the number of nuclei in the ensemble, $\omega_N$ is their Larmor frequency and $A$ is the signal amplitude generated by a single nucleus. 
	In the quantum case $A$ is replaced by $\phi$, the phase acquired from a single nucleus, which is a function of the interaction strength and the interrogation time.
	The uncertainty of the frequency measurement of the signal \eqref{P_classic} scales inversely with the derivative and thus can be written as
	\beq\label{motivation}
	\Delta\omega_N=\frac{C}{N t}=\frac{1}{N t \phi},
	\eeq
	 where the first equality is correct for either classic or quantum sensing with some constant $C$, while the second is correct only for the quantum case.
	 We know, however, that in a noisy environment the optimal precision, achievable by performing Ramsey spectroscopy on each nucleus \cite{QFI,MartinSQL} is
	 \beq\label{Ramsey_limit}
	 \Delta\omega_N\leq\frac{1}{\sqrt{N} t}.
	 \eeq 
	 Comparing Eqs. \eqref{motivation} and \eqref{Ramsey_limit} yields the restriction $\phi\leq\frac{1}{\sqrt{N}}$. Since the phase is a function of the interaction between the sensor and the nuclei, which increases as the sensor is brought into close proximity with the sample, the behavior of the signal must change when $\phi\sim\frac{1}{\sqrt{N}}$. Therefore, there is a critical depth $d_c$, for which the behavior of the signal transforms from classical to quantum (see the right side of Fig. \ref{System}). For common parameters of liquids, an interrogation time of $T_2^{NV}=1 \ \textrm{ms}$ and a fully polarized nuclear spin ensemble, we find that $d_c\sim140\ nm$ (see section 3 for details). For finite polarization, $pol$, the critical distance changes $d_c\sim pol^{2/3}T_2^{2/3}$, such that for $pol=1\%$ it drops to $d_c\sim6.5\ \textrm{nm}$ for the same $T_2$.  
	 We henceforth refer to Eq. \eqref{motivation} as "Heisenberg scaling with $N$" or as the "weak limit" for reasons that will readily become apparent, and to Eq. \eqref{Ramsey_limit} as the Ramsey limit or the standard quantum limit (SQL).       
%Santi says that this discussion is a bit convoluted. Maybe we should consider simplifying.
	
	Here we provide and analyze a nano-NMR protocol and show that for an NV with a given depth, for short measurement times, our protocol achieves a "Heisenberg-like" scaling with N, which is greatly reduced by the weak NV-nucleus coupling, whereas for long measurement times, the scaling changes to the SQL with respect to N and does not depend on the coupling. 
	Moreover, we show that using an adaptive measurement method, for long measurement times, the uncertainty scales effectively, up to a prefactor, as the fundamental Ramsey limit that requires individual addressing of each nucleus.  
	Alternatively, for a given measurement time, the precision of the protocol changes from the weak limit to the SQL for increasingly shallow NVs. In other words, even when using shallow NVs for probing an ensemble with finite polarization in the weak coupling limit - where the NV-nucleus coupling constant multiplied by the measurement time is much smaller than $\pi$ - our protocol achieves an accuracy scaling that does not strictly depend on the coherence time of the NV or the NV-nucleus coupling constant and follows the SQL with respect to N.  
	
	The paper is organized as follows. We start with a simplified model for nano-NMR to introduce our protocol and the main results. Then we explain how this model can be adapted to include more realistic settings and we modify our results accordingly. Finally, we discuss possible modifications to the protocol and provide an overview of the work.          		
	\section{results}
	\subsection{Simplified Model: Constant Coupling}
	Let us start with a simplified model that captures the essentials. The physical system consists of a quantum sensor, taken to be a two level system with energy gap $\omega_0$, and an ensemble of $N$ spin-1/2 nuclei with energy gap $\omega_N$ as described by the Hamiltonian
	\beq\label{H_free}
	H_0=\frac{\omega_0}{2} \sigma_z+\frac{\omega_N}{2} \sum\limits_{i=1}^{N} I_z^i,
	\eeq
	where $\sigma_i/I_i^j$ is the Pauli matrix of the sensor/j-th nucleus in the $i$ direction.
	The NV and the nuclei interact with a constant coupling 
	\beq\label{H_toy}
	H_1=g\sigma_z \sum_{j=1}^{N}I_x^j.
	\eeq
	Our protocol for the estimation of $\omega_N$ starts by initializing the system to the state $
	\ket{\psi_i}=\ket{\uparrow_X}^{S}\otimes\ket{\uparrow_X\cdots\uparrow_X}$
	and the interaction is suppressed. Note, that we explicitly assume that the nuclear spins are completely polarized in the $x$ direction. The propagation in the interaction picture with respect to the sensor's energy gap is
	\beq
	\ket{\psi_t}=\ket{\uparrow_X}^{S}\otimes\left(\cos\left(\frac{\omega_N}{2}t\right)\ket{\uparrow_X}-i\sin\left(\frac{\omega_N}{2}t\right)\ket{\downarrow_X}\right)^{\otimes N}.
	\eeq
	Then $H_1$ is turned on while $H_0$ is turned off, so the propagation by an additional time $\tau$ will result in
	\beq
	\ket{\psi_\tau}=\frac{1}{\sqrt{2}}\left(\ket{\uparrow_Z}^{S}\ket{\psi_+}+\ket{\downarrow_Z}^{S}\ket{\psi_-}\right),
	\eeq
	where $
	\ket{\psi_{\pm}}=\left(\cos\left(\frac{\omega_N}{2}t\right)e^{\mp ig\tau}\ket{\uparrow_X}-i\sin\left(\frac{\omega_N}{2}t\right)e^{\pm ig\tau}\ket{\downarrow_X}\right)^{\otimes N}$.
	 %In practice, the measurement is performed by adding a control driving field on the quantum prob. Then $\omega_N$ is replaced by an effective frequency, as shown in the next section, and the optimal time $t$ can be set close to the coherence time of the nuclei.% 
	The reduced density matrix of the sensor after both steps is
	\beq\label{density}
	\rho_{S}=\frac{1}{2}\begin{pmatrix}1 & \langle\psi_{-}|\psi_{+}\rangle\\
		\langle\psi_{+}|\psi_{-}\rangle & 1
	\end{pmatrix},
	\eeq
	where the off diagonal element is given by
	\beq
	\label{constant}
	\langle\psi_{+}|\psi_{-}\rangle=\left(\cos\left(2g\tau\right)+i\sin\left(2g\tau\right)\cos\left(\omega_N t\right)\right)^{N}.
	\eeq
	\begin{figure}[t]
		\begin{center}
			\includegraphics[width=0.48\textwidth]{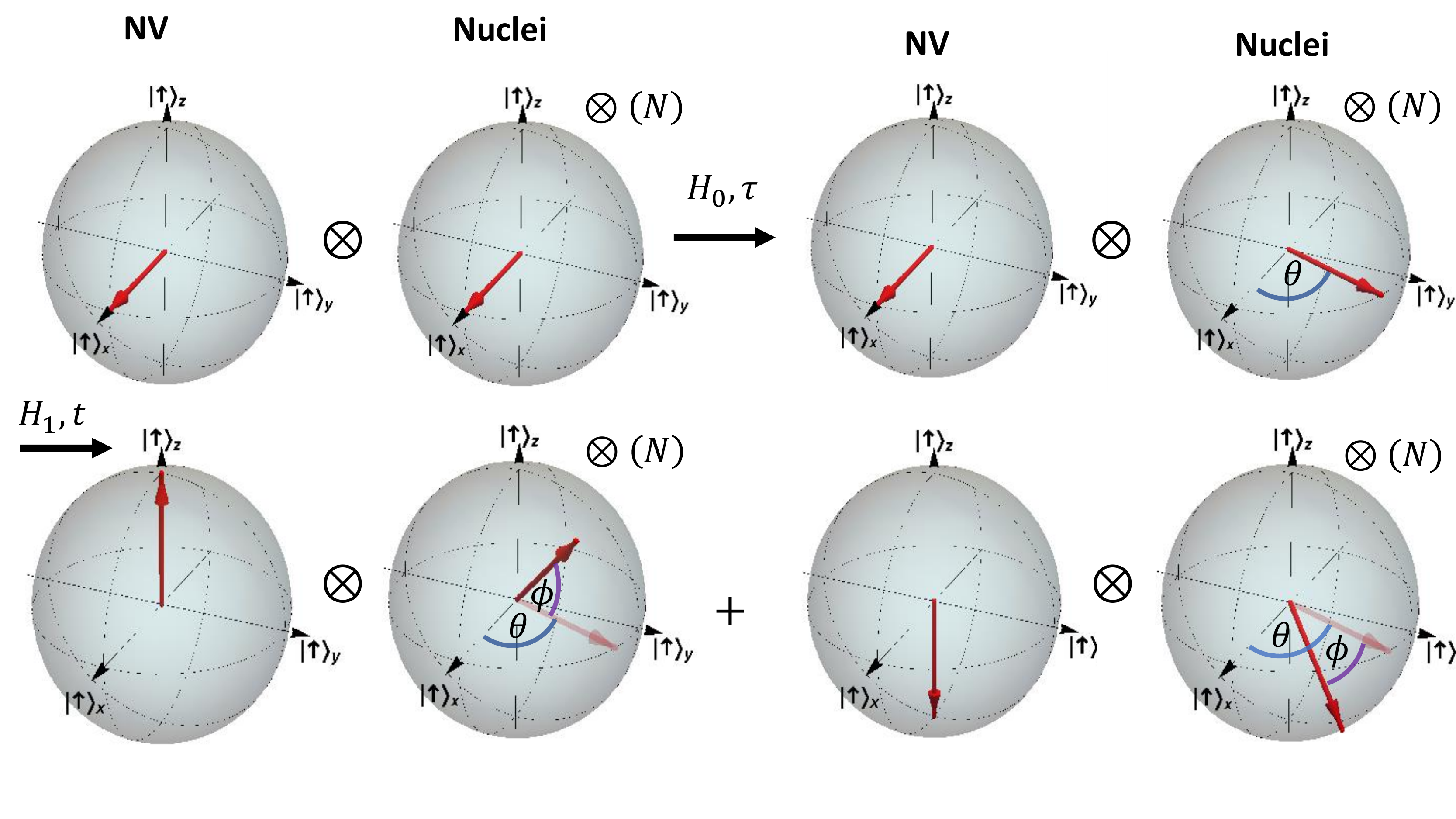}
		\end{center}
		\caption{{\em The protocol ---}  The NV and the nuclear spins are initialized at the $x$ direction. The nuclei are then allowed, using an appropriate pulse sequence, to propagate according to the free Hamiltonian $H_0$ for a time $t$, which results in a rotation by $\theta=\omega_N t$ around the $z$ axis. Then, by changing the external pulse sequence, the system propagates under the interaction Hamiltonian $H_1$ for a duration of $\tau$. This results in a rotation of the nuclei around the $x$ axis by an angle of $\pm\phi=\pm 2 g\tau$ depending on the sensor's state. The sensor will experience an effective dephasing depending on the extent of its entanglement with the ensemble.}
		\label{Dephasing}
	\end{figure}
	To obtain the product it is convenient to write the complex number \eqref{constant} in a polar representation  
	\begin{align}
	&\Phi=N\arctan\left[\tan\left(2 g\tau\right)\cos\left(\omega_N t\right)\right]\approx 2 N g\tau\cos\left(\omega_N t\right)\label{signal_const}\\
	&r=\left( 1-\sin^2\left(2g\tau\right)\sin^2\left(\omega_N t\right) \right) ^{N/2} \approx e^{-2 N \left(g\tau\right)^2\sin^2\left(\omega_N t\right)}\label{decay_const},
	\end{align}
	where the approximations are made using the assumption of weak coupling, $g\tau\ll1$.
	We henceforth refer to the accumulated phase, $\Phi$, as "the signal" since it corresponds to the "classical" signal \eqref{P_classic} with the aforementioned accumulated phase $\phi=2g\tau$ , and we refer to $r$ as the decay.   
	The decay is caused by the entanglement between the NV and nuclear spin ensemble, as illustrated in fig. \ref{Dephasing}. For times $\omega_N t=n \pi$ the dynamics is purely classical because the rotation around the $x$ axis is trivial, whereas for other times the rotation causes the sensor state to be entangled with the collective state of the ensemble. Therefore, the dimensionless quantity characterizing the decay, $N\left(g\tau\right)^2$, is interpreted as the back-action. Because the coupling constant $g$ depends on the distance between the sensor and the sample, the transition from weak to strong back-action is dictated by the NV's depth for a given $\tau$. This transition is the one predicted in the introduction, as it occurs when $\phi=2gt\sim\frac{1}{\sqrt{N}}$.      
	Substituting Eqs. \eqref{signal_const} and \eqref{decay_const} into \eqref{constant},
	\beq\label{constant2}
	\langle\psi_{+}|\psi_{-}\rangle=r e^{i\Phi}\approx e^{-2N\left(g\tau\right)^2\sin^2\left(\omega_N t\right)} e^{i2 N g\tau\cos\left(\omega_N t\right)}.
	\eeq
	The measurable quantity associated with \eqref{constant2} is the probability distribution, 
	\beq\label{prob_y}
	P_{\ket{\uparrow_Y}}=\frac{1}{2}\left(1+ e^{-2 g^2 N \tau ^2 \sin ^2(\omega_N t)} \sin (2 g N \tau  \cos (\omega_N t))\right),
	\eeq
	which is related to eq. \eqref{constant2} by $P_{\ket{\uparrow_Y}}=\frac{1}{2}\left(1-\textrm{Im}\left[\langle\psi_{+}|\psi_{-}\rangle\right]\right)$.
	The function \eqref{prob_y} is the non-approximate form of \eqref{P_classic}. It is drawn on the right hand side of Fig. \ref{System} with $N=3\cdot 10^5$  and $g\tau=5\cdot10^{-5}$  (top) or $g\tau=0.01$  (bottom) which correspond to weak and strong back-action respectively.   
	
	In order to determine the precision of the estimation of $\omega_N$ we use the tools of quantum metrology. The Fisher Information for the discrete probability $\left\{p_i\right\}_{i}$ dependent on a parameter g is defined as $I=\sum_i \frac{\left(\frac{dp_i}{dg}\right)^2}{p_i}$ \cite{Information_theory}. For a quantum system one can optimize over all possible measurement bases. This leads to the definition of Quantum Fisher Information (QFI) \cite{degen2017quantum}. For a density matrix $\rho=\sum_i p_i\left(g\right)\ket{\Psi_i\left(g\right)}\bra{\Psi_i\left(g\right)}$,
	 the QFI about $g$ is given by $\pazocal{I}=\sum_{p_i+p_j\neq0}\frac{2}{p_i+p_j}\left|\bra{\Psi_j}\frac{d\rho}{d g}\ket{\Psi_i}\right|^2$ \cite{QFI}. 
	The precision of any measurement is bounded by the Cramer-Rao bound, $\Delta g\leq\frac{1}{\sqrt{\pazocal{I}}}$. Since this is a tight bound, we use it henceforth to quantify precision.
	 For the density matrix \eqref{density} the QFI can be expressed as the relevant Bures distance \cite{slater1998bures, QFI}:
	\beq\label{formula}
	\pazocal{I}=\frac{\left(\frac{dr}{d\omega_{N}}\right)^{2}}{1-r^{2}}+r^{2}\left(\frac{d\Phi}{d\omega_{N}}\right)^{2},
	\eeq 
	thus, eqs. \eqref{signal_const}, \eqref{decay_const} and \eqref{formula} lead to
	\begin{align}\label{QFI_const}
	\pazocal{I}=\phi^2 N^2 t^2 \sin ^2(\theta ) e^{-\phi^2 N \sin ^2(\theta )}\left[1+\frac{\phi^2 \cos ^2(\theta )}{1-e^{-\phi^2 N \sin ^2(\theta )}}\right],
	\end{align}
	where $\theta=\omega_N t$. 
	In the limit of weak back-action, $N \left(g \tau\right)^2 \ll 1$,
	\begin{align}
	&P_{\ket{\uparrow_Y}}=\frac{1}{2}\left(1+\sin ( N \phi  \cos (\omega_N t))\right),\\
	&\pazocal{I}\approx \phi^2 N^2 t^2 \sin ^2\left(\theta \right)\left[1+\frac{\cos ^2(\theta )}{ N \sin ^2(\theta )}\right].
	\end{align}
	The QFI in this limit is optimal when
	\beq\label{optimal_time_const_low_dec}
	\sin^2\left(\theta_{opt}\right)=1,
	\eeq 
	since the derivative of the signal is then maximal and the decay is negligible. For the optimal time \eqref{optimal_time_const_low_dec}, the QFI \eqref{QFI_const} is
	\beq\label{QFI_toy_low_decay}
	\pazocal{I}=4 \phi^2 N^2 t^2,
	\eeq 
	which corresponds to the uncertainty $
	\Delta\omega=\frac{1}{2 N g \tau t}$ as in \eqref{motivation} with $\phi=2g\tau$. 
	Thus, in the limit of weak back-action, we achieve an uncertainty that scales as $N^{-1}$ in exchange for the large factor $\left(g \tau\right)^{-1}$. 
	
\begin{figure}[t]
		\begin{center}
			\includegraphics[width=0.48\textwidth]{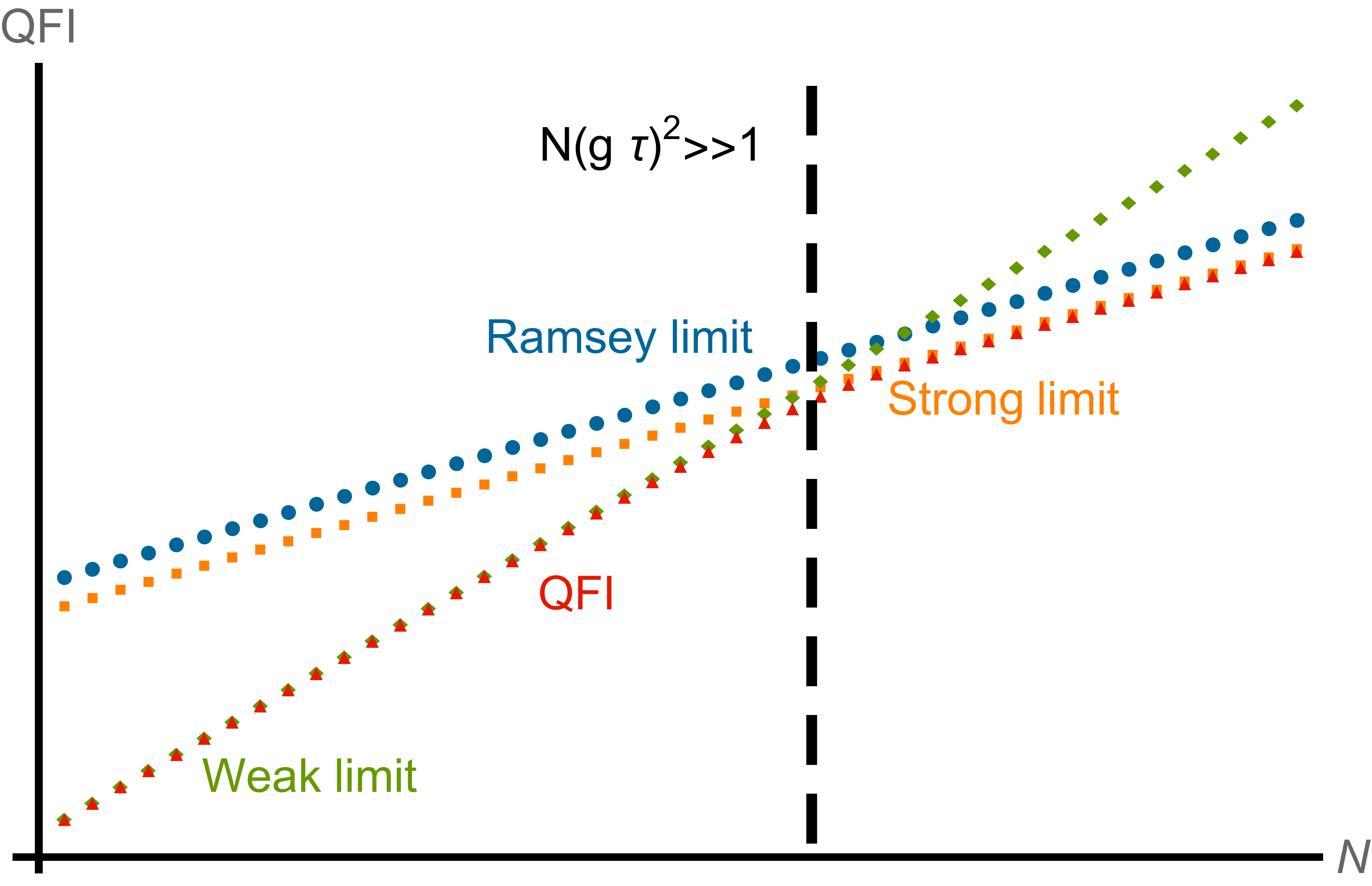}
		\end{center}
		\caption{The maximal value of the QFI \eqref{QFI_const} (red) for $g\tau=0.01$ and $t=1$ as a function of $N$ (log scale). The optimal QFI follows the weak back-action limit (green) given by \eqref{QFI_toy_low_decay} when $N(g \tau)^2\ll1$. As $N$ increases the back-action corrects the scaling until $N(g\tau)>1$, where the QFI approaches the strong limit given by \eqref{QFI_toy}. In the strong back-action regime, the QFI of our scheme is only smaller by a factor of $e^{-1}$ from the optimal QFI, $Nt^2$, achieved by Ramsey spectroscopy on each individual nucleus (blue).}
		\label{QFI_plot}
	\end{figure}

\begin{figure}[t]
		\begin{center}
			\includegraphics[width=0.48\textwidth]{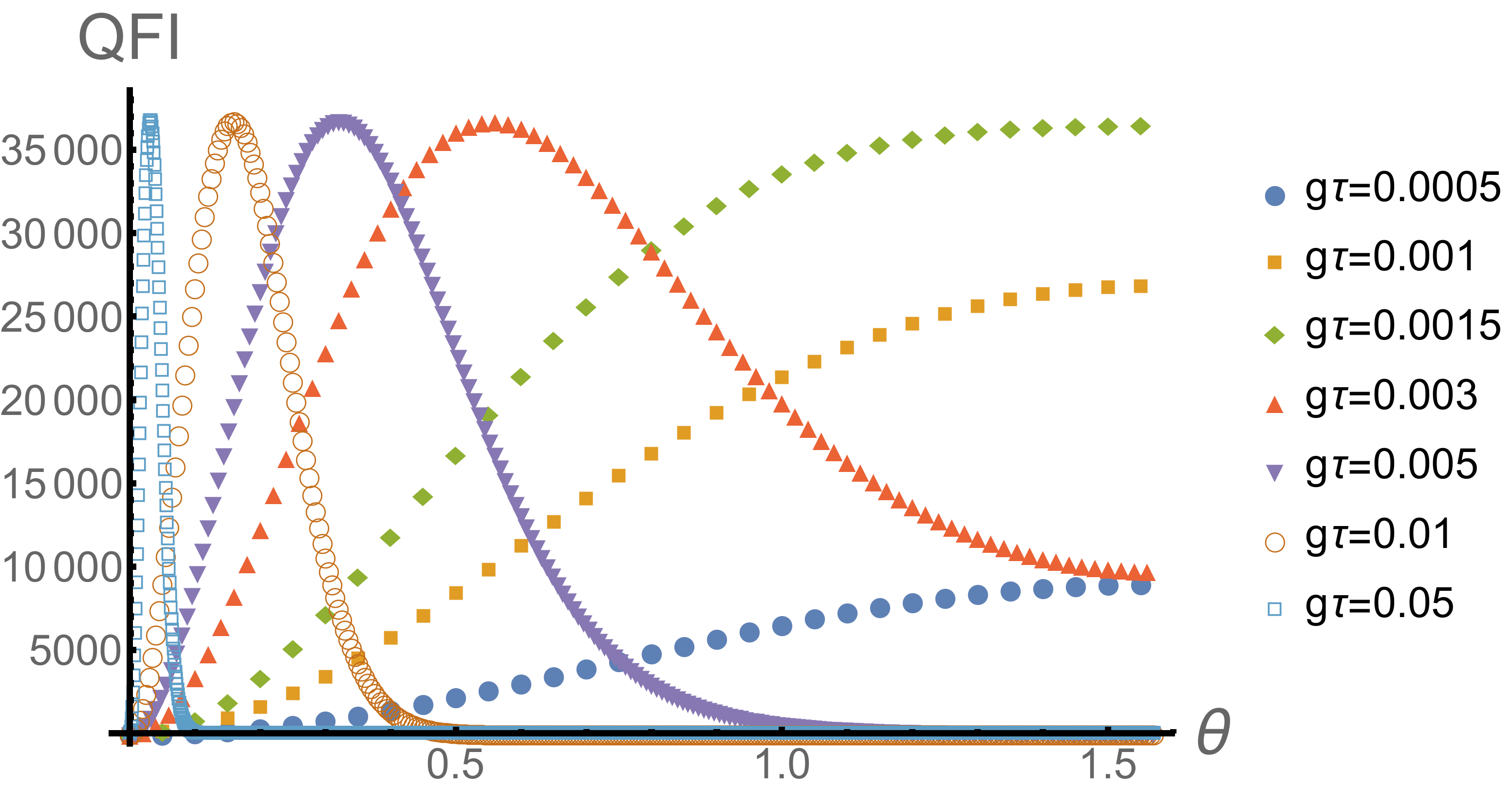}
		\end{center}
		\caption{The QFI \eqref{QFI_const} for $t=1$ and $N=10^5$ for different values of $g\tau$. As the back-action $N(g\tau)^2$ gets stronger the peak of the QFI increases until it reaches an optimum since it no longer depends on the interaction .The increase in QFI comes at the cost of it becoming increasingly narrow. Therefore, to achieve the optimal scaling \eqref{QFI_toy} a good estimate of $\omega_N$ is already required. This can be attained by using an adaptive protocol as explained in \cite{Supp}.}
		\label{QFI_Phi}
	\end{figure}

	Achieving better precision, however, requires $N g^2 \tau^2 \gg 1$, where consequently, the decay starts to affect the quantum sensor.  The optimal time \eqref{optimal_time_const_low_dec}, therefore, must change to account for the decay. The new time is approximately      
	\beq\label{optimal_time_const}
	\sin^2\left(\theta_{opt}\right)=\frac{1}{N\phi^2},
	\eeq 
	for which the QFI \eqref{QFI_const} is 
	\beq\label{QFI_toy}
	\pazocal{I}\approx\frac{N t^2}{ e}
	\eeq
	and corresponds to the uncertainty $
	\Delta\omega= \frac{1}{t}\sqrt{\frac{e}{N}}.$
	The decay, therefore, "corrects" the apparent Heisenberg scaling for high precision. This is a surprising  result, as the uncertainty is limited solely by the coherence time of the nuclei as in \cite{qdyne,NanoNMRQdyne,boss2017quantum,glenn2018high,bucher2018hyperpolarization,gefen2018quantum} and it does not depend on the coupling constant $g$. Moreover, if we could manipulate and read-out each nucleus individually, the optimal QFI would be attained by the Ramsey limit \eqref{Ramsey_limit}. Therefore, up to the numerical factor of $e^{-1}$ our protocol is the optimal one. Figure \ref{QFI_plot} compares the different QFI scalings to our protocol.
	To obtain this QFI we assume optimization of $\theta$ and the measurement basis, which requires knowledge of $\omega_N$.
	 As shown in fig. \ref{QFI_Phi}, the QFI increases with $N(g\tau)^2$, but it also becomes increasingly narrow, such that a more accurate estimation of $\omega_N$ is required to achieve it. This, however, can be resolved by using an adaptive measurement protocol - first a non-optimal measurement is performed to acquire an estimate of $\omega_N$, and the next measurement's duration is optimized according to the previous measurement's outcome. This can be repeated until the optimal precision is attained.
	 An analysis of this model using multiple sensors can be found in \cite{Supp}.
	  
	\subsection{Spatially dependent coupling}\label{spatial}
	In the previous section we presented the simplified model with the interaction Hamiltonian \eqref{H_toy}. This is an approximate form of the dipole-dipole interaction Hamiltonian, which can be written as
	\beq
	H_{DD}=\sum_{k=1}^N\sum_{i,j=\left\{x,y,z\right\}} \tilde{g}_{i,j}^k \sigma_i I_j^k.
	\eeq
	In the interaction picture with respect to sensor's energy gap, after taking the rotating wave approximation, the remaining terms are
	\beq\label{H_DD_no_drive}
	H_{DD}^I\approx \sigma_z \sum_{i=1}^N\left(g_0^i I_z^i+g_1^i I_x^i+g_2^i I_y^i\right).
	\eeq
	If we further assume that the nuclei can be driven in the $x$ direction sufficiently fast compared to the interaction and the entanglement generation rate, we arrive at the Hamiltonian
	\beq\label{H_DD_with_Drive}
	\tilde{H}_1=\sigma_z \sum_{i=1}^N g^i I_x^i.
	\eeq 
	The difference between \eqref{H_DD_with_Drive} and the toy model \eqref{H_toy} is that the coupling differs from one nucleus to the other. It depends, in fact, on the position of the nucleus and, therefore, also on time. Further description of the interaction requires specific realization of the sensor, henceforth we use the NV center. In a spherical coordinate system, where the NV is found at the origin, the NV's magnetization axis coincides with the $z$ axis and the $i-th$ nucleus position is denoted by $\left\{r_i,\theta_i,\varphi_i \right\}$, the coefficients $g_i$ are given by
	\beq
	g^i\left(t\right)\equiv g\left(\bar{r}_i\left(t\right)\right)=-\frac{3}{2} J \left[r_i\left(t\right)\right]^{-3}\sin\left[\theta_i\left(t\right)\right]\cos\left[\theta_i\left(t\right)\right]\cos\left[\varphi_i\left(t\right)\right],
	\eeq   
	 with the physical coupling constant $J=\frac{\mu _0 \hbar\gamma_e\gamma_N}{4\pi}=0.49 \textrm{MHz}\cdot\textrm{nm}^3$, where $\mu _0$ is the vacuum permeability, $\hbar$ is the reduced Plank constant and $\gamma_{e/N}$ are the electronic/nuclear gyro-magnetic ratios.  
	Repeating the derivation of the previous section with the interaction \eqref{H_DD_with_Drive}, eq. \eqref{constant} becomes
	\beq\label{Drive}
	\langle\psi_{+}|\psi_{-}\rangle=\prod_{j=1}^N \left(\cos\left(2G_j\right)+i\sin\left(2G_j\tau\right)\cos\left(\omega_N t\right)\right),
	\eeq
	where $G_j=\int\limits_0^\tau dt g_j\left(t\right)$.
	In the limit of weak coupling, $G_i\ll1$, \eqref{Drive} can be approximated to 
	\beq
	\label{Drive2}
	\langle\psi_{+}|\psi_{-}\rangle=e^{-2\sum_{i=1}^NG_i^2\sin^2\left(\omega_N t\right)}e^{2 i\sum_{i=1}^N G_i\cos\left(\omega_N t\right)}.
	\eeq
	
There is therefore a dependence only on $s_{1}=\underset{i}{\sum}G_{i},s_{2}=\underset{i}{\sum}G_{i}^{2}$ and the QFI reads,	
\begin{equation}\label{QFI_drive}
\pazocal{I}\approx4 t^{2} s_{1}^{2}\sin\left(\theta\right)^{2}\exp\left(-4s_{2}\sin^{2}\left(\theta\right)\right).
\end{equation}
The quantity that determines the behavior of the optimal QFI is $s_2$:
\begin{equation}\label{QFI_drive_opt}	
\pazocal{I}\approx \begin{cases}
 4 t^{2} s_{1}^{2} & s_{2}\ll1,\ \sin^2\left(\theta\right)=1 \\
t^{2}  \frac{1}{e}\frac{s_{1}^{2}}{s_{2}} & s_{2}\gg1,\ \sin^2\left(\theta\right)=\frac{1}{4s_2}
\end{cases}.
\end{equation}
The regime of interest is $s_2\gg1$, where, as we readily show, a modified SQL scaling can be achieved.
Note that $\pazocal{I}\leq t^{2}N/e,$ where equality is attained only for homogeneous couplings.

To obtain the QFI in a nano-NMR scenario, we need therefore to find the relevant $s_{1},s_{2}$. 
The results will of course depend on the parameters of the problem. We denote by $n=\frac{N}{V}$ the nuclei number density, $D$ the diffusion coefficient of the sample, $d$ the depth of the NV and $\alpha$ the NV's tilting angle, measured between the normal to the diamond surface and it's magnetization axis. 
Let us first calculate $s_{1}$ assuming $N\gg1$, 
\beq\label{signal_drive}
	s_{1}=\gamma_e\left<B\right>\tau= n \tau\int d^3r g\left(\bar{r}\right)=-\pi n J \sin\left(2\alpha\right)
	\eeq
While $s_{1}$ is proportional to the average magnetic field, it can be observed that $s_{2}$ goes as the magnetic field auto-correlation,
%$s_{2}$ can be expressed in terms of the magnetic field auto-correlation function:
 \begin{align}\label{decay_drive}
 s_{2}&= \gamma_e^2 \int\limits_{0}^{\tau}dt\int\limits_{0}^{\tau}dt_0\int d^3r d^3r_0 g\left(\bar{r}\right)g\left(\bar{r_0}\right) P\left(\bar{r},\bar{r}_0,t-t_0\right) \\\nonumber
 &= \gamma_e^2 \int\limits_{0}^{\tau}dt\int\limits_{0}^{\tau}dt_0\left<B\left(t\right)B\left(t_0\right)\right>
 \end{align}
where  $\left<\cdot\right>$ is an average over realizations and $P(\bar{r},\bar{r}_0,t)$ is the stationary diffusion propagator from $\bar{r},$ to $\bar{r}_0$ with time difference $t-t_0$.
The correlation decays with a characteristic time of $\tau_D\equiv\frac{d^2}{D}$, which dictates the behavior of $s_2.$
Although the full expression of $s_{2}$ is involved \cite{microfludics}, the asymptotic behavior is rather simple,
%\beq\label{decay_drive_all}
%s_2=\begin{cases}
%	\gamma_e^2 B_{rms}^2\tau^2 & \tau\ll\tau_D\\
%	S\left(\omega=0\right)\tau=\gamma_e^2 B_{rms}^2\tau\tau_D &  \tau\gg\tau_D
%\end{cases}
%\eeq 
\beq\label{decay_drive_all}
s_2=\begin{cases}
	\gamma_e^2 B_{rms}^2\tau^2 & \tau\ll\tau_D\\
	\gamma_e^2 B_{rms}^2\tau\tau_D &  \tau\gg\tau_D
\end{cases},
\eeq 
where  $B_{rms}^2$ is the instantaneous fluctuation in the magnetic field:
\begin{align}\label{angular_2}
	&  \gamma_{e}^{2}  B_{rms}^{2}= \gamma_{e}^{2}  \langle B^{2}\rangle=\int d^{3}r\,g\left(\bar{r}\right)^{2}  \\\nonumber
	&=\pi n J^2\left(\frac{35  -3  \cos \left(4 \alpha \right)}{256 d^3}\right).
\end{align}
The mean field, \eqref{signal_drive}, and $B_{rms}$,  \eqref{angular_2}, can be estimated more explicitly,
\begin{align}
&\gamma_e\left|\left<B\right>\right|\approx\frac{n}{n_{water}}48 \textrm{MHz},\\
&\gamma_e^2B_{rms}^2\approx\frac{n}{n_{water}}\left(44.5 KHz\right)^2\left(\frac{10\textrm{nm}}{d}\right)^3,
\end{align}
where we took $\alpha=54.7^\circ$ and $n_{water}=33 \ \textrm{nm}^{-3}$ as the water number density. We use these quantities for the remainder of the article in our quantitative estimates.

The validity condition of the second order approximation in $G_i$ can be put into physical terms with the definitions \eqref{signal_drive} and \eqref{angular_2} ; e.g. Eqs. \eqref{Drive2} and \eqref{QFI_drive} are valid when  $\frac{B_{rms}}{\left|\left<B\right>\right|}\ll1$, which means that the polarization should be larger than the statistic polarization. Hence, for increasingly shallow NVs, higher orders should be taken into account (see discussion in \cite{Supp}).

%We can now insert $s_{1}$ (eq. \eqref{signal_drive}) and $s_{2}$ (eq. \eqref{decay_drive_all}) in \eqref{QFI_drive_opt} to get the QFI.
We are now fully equipped to calculate the optimal QFI by substituting $s_1$ and $s_2$ (Eqs. \eqref{signal_drive} and \eqref{decay_drive_all} respectively) into \eqref{QFI_drive_opt}.
%	\begin{equation}\label{QFI_drive2}
% 	\pazocal{I}= \begin{cases} 
% 	4 \gamma_e^2\left<B\right>^2\tau^2t^2 \sin ^2(\theta ) e^{-4 \gamma _e^2 B_{rms}^2 \tau ^2  \sin ^2(\theta )} & \tau\ll\tau_D\\
% 	4 \gamma_e^2\left<B\right>^2\tau^2t^2 \sin ^2(\theta ) e^{-4 \gamma _e^2 B_{rms}^2 \tau \tau_D  \sin ^2(\theta )}  &  \tau\gg\tau_D
% 	\end{cases}
%	\end{equation}
	In the weak back-action regime, $s_2\ll1$, the optimal QFI reads:
	\beq\label{QFI_drive_short_times_low_decay}
	\pazocal{I}=4 \gamma_e^2 \left<B\right>^2 \tau^2 t^2.
	\eeq 
	For strong back-action the optimal QFI depends explicitly on $\tau_D$,
	\beq
	\pazocal{I}\approx \begin{cases}\label{QFI_drive_large_decay}
		\frac{\left<B\right>^2}{eB_{rms}^2}t^2
		\approx17.5\frac{nd^3}{e}t^2 & \tau\ll\tau_D \\
	\frac{\left<B\right>^2 }{eB_{rms}^2}\frac{\tau}{\tau_D}t^2\approx17.5\frac{nd^3}{e}\frac{\tau}{\tau_D}t^2 & \tau\gg\tau_D
	\end{cases},
	\eeq
	where the optimal $\theta$ is given by
	\beq\label{opt_theta_drive}
	\sin^2\left(\theta\right)\approx \begin{cases}
	\frac{1}{4 \left(\gamma_e B_{rms} \tau\right)^2} & \tau\ll\tau_D \\
	\frac{1}{4 \left(\gamma_e B_{rms} \right)^2\tau\tau_D} & \tau\gg\tau_D
	\end{cases}.
	\eeq
Note that the transition from \eqref{QFI_drive_short_times_low_decay} to \eqref{QFI_drive_large_decay} occurs when $\gamma_eB_{rms}T_2^{NV}=1$, which defines the critical depth $d_c$ in terms of the physical parameters.

The result \eqref{QFI_drive_short_times_low_decay} and the short times limit of \eqref{QFI_drive_large_decay} are similar to those of the simplified model, with the difference that due to the dipolar interaction and the geometry, there is an additional dependence on the NV's tilting angle. In \eqref{QFI_drive_large_decay} this is translated into a change in the total number of particles in \eqref{QFI_toy} by an effective number $N\approx17.5nd^3$, which is proportional to the number of particles in the effective interaction region.
	
The long times regime of Eq. \eqref{QFI_drive_large_decay} may look puzzling since for large enough $\tau$ we can get an arbitrarily large QFI, which seems paradoxical.
 This arbitrarily large QFI is due to the fact that we consider an infinite sample volume; i.e. an infinite amount of nuclear spins.
 Restricting ourselves to a finite volume, $V$, with $N$ nuclei yields the restriction $\frac{\left<B\right>^2 }{eB_{rms}^2}\frac{\tau}{\tau_D}t^2\leq N t^2$ by imposing the SQL. This limitation on $\tau$ can be taken to be $\tau\ll\tau_V$ \cite{Supp}, where $\tau_V=V^{2/3} /D$ is the volumetric diffusion time; i.e., the characteristic time it takes for a particle to move from one of the volume's boundaries to another. For sufficiently long times, $\tau\gg\tau_V$, the QFI scaling changes to 
	\beq\label{spatial_SQL}
	\pazocal{I}\approx\frac{n V t^2}{e}=\frac{N}{e}t^2.
	\eeq
	This is the equivalent of \eqref{QFI_toy}. In the simplified model the sensor is coupled equally to all the nuclei from the start; therefore, the QFI scales with the total number of nuclei, while in reality the dipolar interaction creates an effective cut off, such that in short times the QFI scales as $nd^3$ and only after a long time, $\tau\gg\tau_V$, when all the nuclei have passed through the interaction region, does the QFI scale with the total number of nuclei. As in Eq. \eqref{QFI_toy}, in this model the QFI is not limited by the coherence time of the NV, $T_2^{NV}$, but only by the coherence time of the nuclei $T_2^{N}$ which is usually longer by orders of magnitude.
	
	\subsection{Undriven nuclei}
In the previous section we assumed that we can drive the nuclei sufficiently fast to achieve the approximate dynamics given by \eqref{H_DD_with_Drive}. In some experimental settings this is unfeasible for practical reasons. Hence, in what follows, we drop this assumption.
The approximate dipolar Hamiltonian \eqref{H_DD_no_drive} is rewritten as
\beq
H_1=\sigma_z\sum_{i=1}^N \left(g_0^i I_z^i+g_1^i  I_+^i +g_2^i  I_-^i\right).
\eeq
The NV is then driven with $\pi$ pulses every time $\tau_p$ so that the effective pulse frequency $\omega_p=\frac{\pi}{\tau_p}$ is close to $\omega_N$. This yields the effective Hamiltonian \cite{Supp}
\beq\label{H_eff_pulses1}
H_1\approx\frac{2g^i_{\pm}}{\pi} \sigma_z\left[  I_x^i\sin\left(\varphi_i\right) - I_y^i\cos\left(\varphi_i\right)\right],
\eeq
and the remaining free Hamiltonian
\beq
H_0=\frac{\delta\omega}{2}\sum_{i=1}^NI_z^i
\eeq
where $\delta\omega=\omega_N-\omega_p$ and $g_{\pm}^i=-\frac{3}{2} J \left[r_i\left(t\right)\right]^{-3}\sin\left[\theta_i\left(t\right)\right]\cos\left[\theta_i\left(t\right)\right]$.
Following the same protocol as before \cite{Supp} the signal is
\begin{align}\label{signal_no_drive}
\Phi=\frac{2}{\pi}\gamma_e\left<B\right>\tau\sin\left(\delta\omega t\right),
\end{align}
where the average magnetic field is as in the driven case, \eqref{signal_drive}. The decay, however, changes to
\begin{align}\label{decay_no_drive}
r&=\exp\left\{-\gamma_e^2\left[\left(B_{rms}^{UDQ}\right)^2\cos^2\left(\delta\omega t\right)+\left(B_{rms}^{UDC}\right)^2\right]\tau^2\right\},
\end{align}
for times $\tau\ll\tau_D$, where 
\begin{align}\label{angular_3}
&\left(B_{rms}^{UDQ}\right)^2=\frac{4 }{\pi^2}B_{rms}^2\left(\frac{2 \sin ^2(\alpha ) (-13 \cos (2 \alpha )+1)}{13 \cos (4 \alpha )+51}\right), \\\label{angular_4}
&\left(B_{rms}^{UDC}\right)^2=\frac{4 }{\pi^2}B_{rms}^2\left(\frac{28 \cos (2 \alpha )+13 \cos (4 \alpha )+87}{52 \cos (4 \alpha )+204}\right),
\end{align}
are the quantum and classical contributions to the instantaneous fluctuation of the magnetic field and $B_{rms}$ is given by \eqref{angular_2}.

Comparing Eqs. \eqref{signal_no_drive} and \eqref{decay_no_drive} to the results of the previous section \eqref{signal_drive} and \eqref{decay_drive_all}, note that the signal is the same upto a prefactor of order $1$ and a constant known phase. The decay, however changes, so it is non-zero for any given time $t$. This is expected, since previously the interaction caused all the nuclei to rotate around the same axis, which resulted in an entanglement induced decay. Without the external drive, the interaction causes each nuclear spin to rotate around a different axis in the xy plane of the Bloch sphere, which induces additional classical dephasing (see supplementary fig. S2). It is worth pointing out that the quantum contribution \eqref{angular_3} can be negative, but the total $B_{rms}$ is always real and positive.
When the decay is small, $\gamma_e B_{rms}\tau\ll1$, we retrieve the result \eqref{QFI_drive_short_times_low_decay} up to a prefactor $\sim1$, since it only depends on the signal. 

If the decay is dominant, the classical and quantum decay compete when optimizing the QFI. On the one hand the strong back-action regime requires $\gamma_e\left|B^{UDQ}_{rms}\right|\tau>1$, which implies that $\tau$ has to be large enough. On the other hand, the classical dephasing causes an exponential decrease in the QFI, $e^{-\left(\gamma_eB_{rms}^{UDC}\tau\right)^2}$, that depends only on $\tau$. To limit this effect we require $\gamma_eB_{rms}^{UDC}\tau<1$. Hence, when the nuclei are undriven fine tuning of $\tau$ is required in order to optimize these competing processes. The optimal protocol will no longer be universal and will depend on the physical parameters.    

First, we assume strong back-action, in order to derive the analog protocol to the previous sections, and to emphasis our last statement regarding the decay. 
The optimal $t$ satisfies
\beq\label{opt_cond_no_drive}
\cos^2\left(\delta\omega t\right)=\frac{1}{\gamma_e^2\left|\left(B_{rms}^{UDQ}\right)^2\right|\tau^2},
\eeq
for which the QFI is
\beq\label{QFI_no_Drive}
\pazocal{I}\approx\frac{\left<B\right>^2 t^2}{e^{\textrm{sgn}\left((B_{rms}^{UDQ})^2\right)} \left|\left(B_{rms}^{UDQ}\right)^2\right|} e^{-\gamma_e^2\left(B_{rms}^{UDC}\right)^2\tau^2}.
\eeq
As aforementioned, the QFI, \eqref{QFI_no_Drive}, has a reduction by an exponential factor compared to the previous result \eqref{QFI_drive_large_decay}, arising from the classical dephasing. 
As long as $\left|\left(B_{rms}^{UDC}\right)^2\right|\leq\left|\left(B_{rms}^{UDQ}\right)^2\right|$ we can satisfy the strong back-action requirement, while limiting the classical decay by taking $\tau^2=\frac{1}{\gamma_e^2\left|\left(B_{rms}^{UDC}\right)^2\right|}$. This results in a reduction by a reasonable factor of $e^{-1}$. This inequality is a function of the NV's tilting angle alone, and is correct for $\alpha\ge\frac{\pi}{3}$. Since the NV's natural tilting angle is $\alpha\sim0.32\pi$ this is approximately the case, where the slight deviation will lead to further reduction of the QFI by a small factor. 
Another issue with this strategy is that it is not guaranteed that $\tau=\frac{1}{\gamma_eB_{rms}^{UDC}}\leq T_2^{NV}$, and indeed for a typical number density and an NV depth of $d=140 \ \textrm{nm}$, the time required is $\tau\approx3\ \textrm{ms}> T_2^{NV}$.    

Obtaining other scalings is possible by optimizing $t$ and $\tau$, depending  on the parameters; for example, by choosing $\cos^2\theta=1,\ \tau^2=\frac{1}{\gamma_e^2\left[\left(B_{rms}^{UDQ}\right)^2+\left(B_{rms}^{UDC}\right)^2\right]}$ we achieve 
\beq
\pazocal{I}=\frac{ \left<B\right>^2}{e B_{rms}^2} t^2,
\eeq
which is the same result as in \eqref{QFI_drive_large_decay}. Note that this is the case of "critical back-action", when $\gamma_eB_{rms}\tau=1$, therefore smaller $\tau$ will lead to the weak back-action limit and larger $\tau$ will cause an exponential decrease.
Since the same typical parameters yield $\tau\approx 2 \ \textrm{ms}$, the lack of external drive reduces Eq. \eqref{QFI_drive_large_decay} by a factor of $e^{-1}$. 
The results for $\tau\gg\tau_D$ can be derived by using reasoning similar to the previous section. However, the long times limit of \eqref{QFI_drive_large_decay} and \eqref{spatial_SQL} will typically no longer be achievable because the optimal $\tau$ will be much longer than $T_2^{NV} $.
	
	\subsection{Finite polarization}
	We now consider the more general case, where the spin's polarization is finite. We assume that the spins are statistically polarized, such that the initial density matrix of the system is
	\beq
	\rho_i=\ket{\uparrow_X}\bra{\uparrow_X}_{NV}\left(p\ket{\uparrow_X}\bra{\uparrow_X}+\left(1-p\right)\ket{\downarrow_X}\bra{\downarrow_X}\right)^{\otimes N}.
	\eeq  
	For the interaction Hamiltonian \eqref{H_DD_with_Drive}, following our protocol, the individual spin's signal and decay (in the limit $\tau\ll\tau_D$) are 
	\begin{align}
	&\Phi=2\left<B\right>\left(2p-1\right)\cos\left(\omega_N t\right),\\
	&r=e^{-2\gamma_e^2B_{rms}^2\tau^2\left[4p\left(1-p\right)+\left(2p-1\right)^2\sin^2\left(\omega_N t\right)\right]},
	\end{align}
	where $\left<B\right>$ and $B_{rms}^2$ are given by \eqref{signal_drive} and \eqref{angular_2}.
	The signal, therefore is the same as the one of a fully polarized ensemble, \eqref{signal_drive} with $\left<B\right>\rightarrow pol\cdot\left<B\right>$, where we denote the polarization by $pol=2p-1$. The decay, on the other hand, consists of two terms: one that can be associated with the polarized dynamics and the other with the unpolarized dynamics. The former, is the same term as the in the fully polarized case \eqref{decay_drive_all}, multiplied by the polarization squared, $(2p-1)^2$. The latter is the constant term which is proportional to the Bernoulli variance and stems from classical dephasing. 
	For a small decay the QFI will then be
	\beq
	\pazocal{I}=4\gamma_e^2\left<B\right>^2\cdot pol^2\tau^2 t^2,
	\eeq
	while for a large decay, as in the previous section, an optimization over $t$ and $\tau$ is required according to the physical parameters. The two possibilities presented in the last section translate here to 
	\begin{align}
	&\tau_{1}^2=\frac{1}{4 \gamma_2^2 B_{rms}^2},\ \sin\left(\theta_1\right)=1\\
	&\tau_{2}^2=\frac{1}{4 B_{rms}^2 pol^2\sin^2\left(\theta_2\right)}
	\end{align}
	which yield
	\begin{align}
	&\pazocal{I}_{1}=pol^2\frac{ \left<B\right>^2}{e B_{rms}^2} t^2,\\
	&\pazocal{I}_{2}=\frac{\left<B\right>^2 t^2 }{e B_{rms}^2}e^{-2\gamma_e^2  B_{rms}^2 \left(1-pol^2\right) \tau ^2 },
	\end{align}
	respectively. 
	\section{Conclusion}
	We provide a protocol for nano-NMR that achieves the SQL up to a prefactor in the strong back-action regime. Moreover, the uncertainty does not depend on the coherence time of the sensor or the probe-nuclear coupling. Our analysis shows that the full SQL, with respect to the number of nuclei, requires a fully polarized sample, an ability to manipulate the nucleus-sensor interaction via a sufficient external drive, and interrogation times longer than the characteristic volumetric diffusion time. Deviation from these standards results in a reduced effective number of particles, which is a function of the aforementioned parameters and the NV's tilting angle. These difficulties, however, can be mitigated by using multiple sensors.  
	These features make it highly applicable for sensing very small samples, which is the ultimate goal of the field.
	\section*{Acknowledgments}
	A. R. acknowledges the support of ERC grant QRES, project No. 770929, grant agreement No 667192(Hyperdiamond), and the ASTERIQS and DiaPol projects.
	
	%\section{Author contributions}
	%A.R. conceived the idea. T.G. and L.O. analyzed the simplified model and devised the protocol. T.G. found the optimal measurement basis. D.C. made the modifications to the protocol for spatially dependent coupling, undriven nuclei and partially polarized ensemble. T.G. and D.C. made estimations of the QFI for multiple quantum probes. D.C. wrote the manuscript with the assistance of L.O., T.G. and A.R. 
	\bibliography{summary_4_arxiv_version}
\end{document}

\title[Short Title]{Supplementary material of achieving the SQL in Quantum NMR spectroscopy without single nucleus addressing}
	\maketitle
	
	\setcounter{figure}{0}
	\renewcommand{\thefigure}{S\arabic{figure}}%

	\section{Simplified model analysis}\label{Simplified}
	The system consists of $N$ spin-$1/2$ nuclei with energy gap $\omega_N$ and a quantum sensor - a spin-$1/2$ particle with energy gap $\omega_0$. The current natural choice for a sensor is an NV center, but the analysis is not limited to a specific realization. The free Hamiltonian of the system is
	\beq\label{const1}
	H_0=\frac{\omega_0}{2} \sigma_z+\frac{\omega_N}{2} \sum\limits_{i=1}^{N} I_z^i,
	\eeq
	where $\sigma_j/I_j^k$ is the Pauli operator in the $j$ direction of the sensor/k-th nucleus. Our goal is to estimate $\omega_N$, when we can only read-out the NV's state and assuming the interaction Hamiltonian
	\beq\label{const2}
	H_1=g\sigma_z\sum_{i=1}^N I_x^i,
	\eeq
	with some constant $g$, which is a simplified version of the dipole-dipole interaction. We propose a new protocol as follows. First initialize the system to the state
	\beq\label{const3}
	\ket{\psi_0}=\ket{\uparrow_X}^{S}\otimes\ket{\uparrow_X\cdots\uparrow_X},
	\eeq
	where superscript $S$ stands for "sensor" and the other registers are the $N$ nuclei.
	The next step is to suppress the dipolar interaction, by applying an appropriate external drive, and propagate the state of system, \eqref{const3}, according to \eqref{const1} in the interaction picture with respect to the sensor's energy gap for a time $t$,
	\beq\label{const4}
	\ket{\psi_t}=\frac{1}{2^{N/2}}\ket{\uparrow_X}^{S}\otimes\left(e^{-i\frac{\omega_N}{2} t}\ket{\uparrow_Z}+e^{i\frac{\omega_N}{2} t}\ket{\downarrow_Z}\right)^{\otimes N}=\ket{\uparrow_X}^{S}\otimes\left(\cos\left(\frac{\omega_N}{2}t\right)\ket{\uparrow_X}-i\sin\left(\frac{\omega_N}{2}t\right)\ket{\downarrow_X}\right)^{\otimes N}.
	\eeq
	Now by changing the drive, we can turn on $H_1$ and turn off $H_0$.
	The propagation of  \eqref{const4} according to \eqref{const2} by time $\tau$ results in
	\begin{align}\label{const5}
	\ket{\psi_\tau}=e^{-iH_1 \tau}\ket{\psi_t}&=\frac{1}{\sqrt{2}}\ket{\uparrow_Z}^{S}\otimes\left(\cos\left(\frac{\omega_N}{2}t\right)e^{-ig\tau}\ket{\uparrow_X}-i\sin\left(\frac{\omega_N}{2}t\right)e^{ig\tau}\ket{\downarrow_X}\right)^{\otimes N}\\\nonumber
	&+\frac{1}{\sqrt{2}}\ket{\downarrow_Z}^{S}\otimes\left(\cos\left(\frac{\omega_N}{2}t\right)e^{ig\tau}\ket{\uparrow_X}-i\sin\left(\frac{\omega_N}{2}t\right)e^{-ig\tau}\ket{\downarrow_X}\right)^{\otimes N}
	\end{align}
	The reduced density matrix of the sensor after this step is
	\beq
	\rho_{S}=\frac{1}{2}\begin{pmatrix}1 & \langle\psi_{2}|\psi_{1}\rangle\\
		\langle\psi_{1}|\psi_{2}\rangle & 1
	\end{pmatrix},
	\eeq
	 where
	 \begin{align}
	 &\ket{\psi_1}=\left(\cos\left(\frac{\omega_N}{2}t\right)e^{-ig\tau}\ket{\uparrow_X}-i\sin\left(\frac{\omega_N}{2}t\right)e^{ig\tau}\ket{\downarrow_X}\right)^{\otimes N},\\
	 &\ket{\psi_2}=\left(\cos\left(\frac{\omega_N}{2}t\right)e^{ig\tau}\ket{\uparrow_X}-i\sin\left(\frac{\omega_N}{2}t\right)e^{-ig\tau}\ket{\downarrow_X}\right)^{\otimes N}.
	 \end{align}
	 The off diagonal element is explicitly,
	 \beq
	 \label{constant}
	 \langle\psi_{1}|\psi_{2}\rangle=\prod_{i=1}^{N}\left(\cos^2\left(\frac{\omega_N}{2}t\right)e^{2ig\tau}+\sin^2\left(\frac{\omega_N}{2}t\right)e^{-2ig\tau}\right)=\left(\cos\left(2g\tau\right)+i\sin\left(2g\tau\right)\cos\left(\omega_N t\right)\right)^{N}
	 \eeq
	 We can rewrite \eqref{constant} in polar representation, $\langle\psi_{1}|\psi_{2}\rangle=r e^{i\Phi}$, with
	 \begin{align}\label{const6}
	 &\Phi=N\arctan\left[\tan\left(2 g\tau\right)\cos\left(\omega_N t\right)\right],\\\label{const7}
	 &r=\left(\cos^2\left(2g\tau\right)+\sin^2\left(2g\tau\right)\cos^2\left(\omega_N t\right)\right)^{N/2}=\left(1-\sin^2\left(2g\tau\right)\sin^2\left(\omega_N t\right)\right)^{N/2}.
	 \end{align}
	 In the weak coupling limit, $g\tau\ll1$, eqs. \eqref{const6} and \eqref{const7} can be approximated by
	 \begin{align}\label{const8}
	 &\Phi\approx 2 N g\tau\cos\left(\omega_N t\right)\\\label{const9}
	 &r\approx\left(1-\left(2g\tau\right)^2\sin^2\left(\omega_N t\right)\right)^{N/2}\approx e^{-2N\left(g\tau\right)^2\sin^2\left(\omega_N t\right)}.
	 \end{align}
	 The off-diagonal elements in this limit are therefore,
	 \beq\label{const10}
	 \langle\psi_{1}|\psi_{2}\rangle=r e^{i\Phi}\approx e^{-2N\left(g\tau\right)^2\sin^2\left(\omega_N t\right)} e^{i2 N g\tau\cos\left(\omega_N t\right)}
	 \eeq
	 The QFI is given by the Bures distance \cite{slater1998bures, QFI}
	 \beq\label{Bures}
	 \pazocal{I}=I_{r}+I_{\Phi}=\frac{\left(\frac{dr}{d\omega_N}\right)^2}{1-r^2}+r^2\left(\frac{d\Phi}{d\omega_N}\right)^2.
	 \eeq
	 
The first term, $I_{r}$, is the information obtained from the change in 
the eignvalues, and the second term, $I_{\Phi}$, is the information obtained from the rotation, i.e. the change in the eigenstates.
Let us analyze each one of these terms (denoting $\theta=\omega_{N} t$).  $I_{r}$ reads:
\begin{equation}
I_{r}  =  N^{2}t^{2}\cos^2\left(\theta\right)\sin^2\left(\theta\right)\sin^4\left(2g\tau\right)\left(1-\sin^2\left(\theta\right)\sin^2\left(2g\tau\right)\right)^{N-2}\frac{1}{1-\left(1-\sin^2\left(\theta\right)\sin^2\left(2g\tau\right)\right)^{N}}.
\end{equation}
Optimizing this term over $\theta$ yields (optimum is obtained at $\theta=0$) $Nt^{2}\sin\left(2g\tau\right)^{2}.$
Hence given a small $g\tau$ this expression is bounded by $\approx4N t^2 \left(g\tau\right)^{2}.$

Let us move on to $I_{\Phi}$, the information due to rotation:
\begin{equation}
I_{\Phi}=N^{2}t^{2}  \cos^2\left(2g\tau\right)   \sin^2\left(2g\tau\right)  \sin^2\left(\theta\right)  \left(1-\sin^2\left(2g\tau\right)\sin^2\left(\theta\right)\right)^{N-1}.
\end{equation}      	   	       	   	       	   	       	   	 
Optimizing this term over $\theta$ yields (optimum is obtained at $\sin^2\left(\theta\right)=\frac{1}{\left(N\right)\sin^2\left(2g\tau\right)}$)
\begin{equation}
I_{\Phi}= Nt^{2}\cos^2\left(2g\tau\right)\left(1-\frac{1}{N}\right)^{N-1}\approx\frac{N}{e}t^{2}\cos^2\left(2g\tau\right).	
\end{equation}  
and in the limit of $g \tau \ll 1$ it is
\beq\label{toy_optimal_strong} 
I_{\Phi}=\frac{N}{e} t^2.
\eeq 
Note that in order to get this information we need to tune $\theta$ such that
$\sin^2\left(\theta\right)=\frac{1}{\left(N\right)\sin^2\left(2g\tau\right)},$ hence we must have $2 N \left( g\tau \right)^{2}>1$.
For $2 N \left( g\tau \right)^{2}<1,$ the optimal $\theta$ is $\pi/2$ and given this optimization the expression behaves as 
\beq\label{toy_optimal_weak}
\pazocal{I}\approx4 N^2 t^2 \left( g \tau  \right)^2. 
\eeq
We can therefore conclude that in the relevant limit, i.e. $g\tau\ll1, N \gg 1,$ $I_{r}$ (information obtained from the length of the Bloch vector) is negligible compared 
to $I_{\Phi}$ (information obtained from rotation of the Bloch vector) and thus $\pazocal{I} \approx I_{\Phi}$. This can be summarized as follows:    
	 
\begin{equation}	
\begin{array}{ccc}
 & I_{r} & I_{\Phi}\\
N\left(g\tau\right)^{2}>1/2 & \,\,4Nt^{2}\left(g\tau\right)^{2} & \frac{N}{e}t^{2}\\
N\left(g\tau\right)^{2}<1/2 & \,\,4Nt^{2}\left(g\tau\right)^{2} & \;4N^{2}t^{2}\left(g\tau\right)^{2}
\end{array}\;\;\Longrightarrow\begin{array}{c}
\pazocal{I}\\
\frac{N}{e}t^{2}\\
4N^{2}t^{2}\left(g\tau\right)^{2}
\end{array}
 \end{equation} 
 and the optimal values $\theta$ that correspond to the optimal $I_{\Phi}$:
\begin{equation} 
\begin{array}{cc}
 & \theta\\
N\left(g\tau\right)^{2}>1/2 & \;\pi k+ \arcsin\left(\frac{1}{\sqrt{N}2g\tau}\right)\\
N\left(g\tau\right)^{2}<1/2 & \;\pi/2
\end{array}	 	 
\end{equation}
where $k\in\mathbb{Z}$.
We would like to point that $I\approx I_{\Phi}$ only in the limit of $g \tau \ll 1,$ for larger values of $g \tau$ this is no longer the case. 	 	 

\begin{figure}[h]
	\begin{center}
		\includegraphics[width=0.6\textwidth]{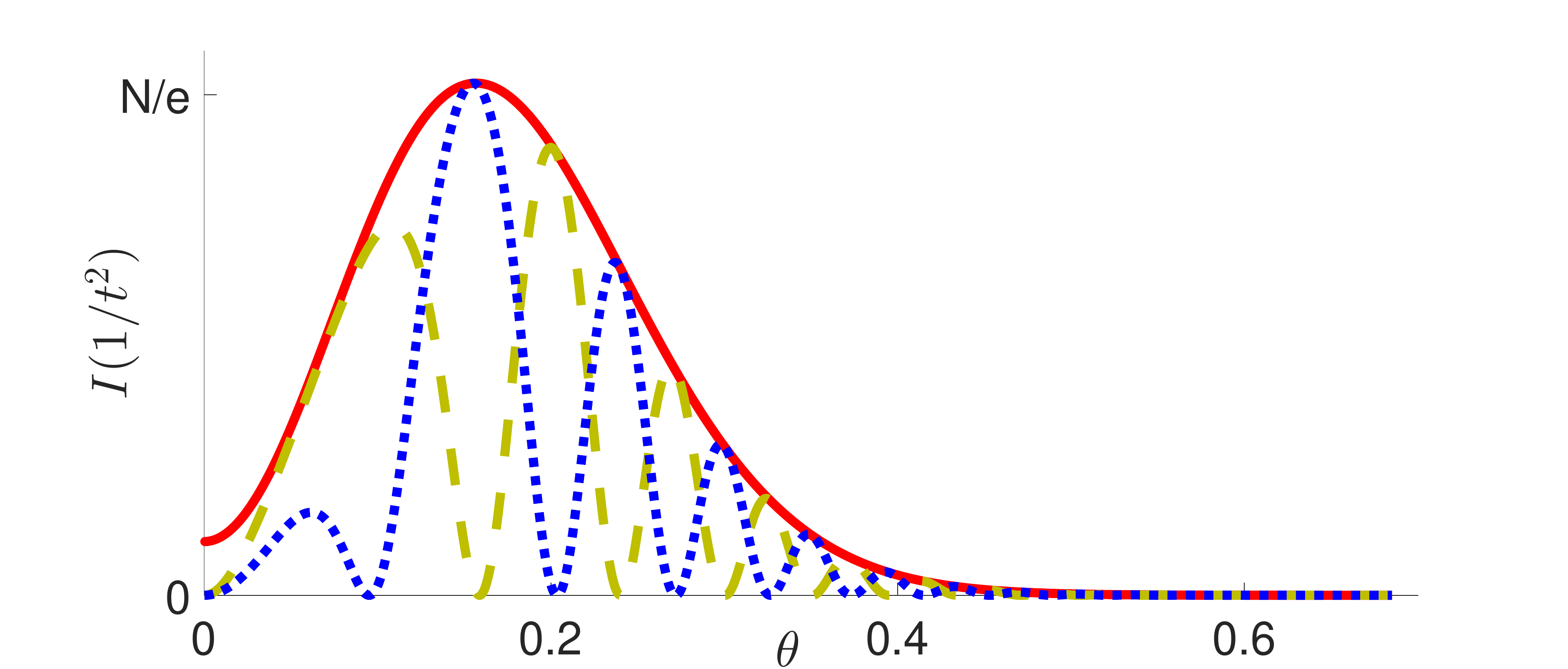}
	\end{center}
	\caption{ Comparison between the FI with different measurements and the QFI (red, solid line). 
	The optimal measurement basis attains the QFI for $\theta_\text{opt}$ (blue, dotted line), 
	a different measurement basis (green, dashed line) leads to a worse performance. }
	\label{bases}
\end{figure}

We obtained that for $N\left(g\tau\right)^{2}\gg1,$ $\pazocal{I}=\frac{N}{e}t^{2},$ hence a QFI that is comparable to the ultimate limit.
It remains to find the measurement basis that saturates the QFI. We can do that by finding the eigenbasis of the symmetric-logarithmic derivative,
however in this case, since almost all the information is in $\Phi$ we can find a simple approximation.
Given a measurement basis that creates an angle of $\alpha$ (in $X-Y$ plane of the Bloch sphere) with $\sigma_{x}$ the probabilities read:   	 
\begin{equation}	 
p_{\pm}=0.5\left(1\pm r\right)\cos^{2}\left(\frac{\alpha-\Phi}{2}\right)+0.5\left(1\mp r\right)\sin^{2}\left(\frac{\alpha-\Phi}{2}\right)=0.5\pm0.5r\cos\left(\alpha-\Phi\right).
\end{equation}	  
Hence the FI given this measurement basis is $I=\sin^{2}\left(\alpha-\Phi\right)\frac{1}{1-r^{2}\cos^{2}\left(\alpha-\Phi\right)}r^{2}\left(\frac{d\Phi}{d\omega_{N}}\right)^{2},$
which attains its maximum ($r^{2}\left(\frac{d\Phi}{d\omega_{N}}\right)^{2}$) at $\alpha=\frac{\pi}{2}+\Phi.$ 
namely the optimal operator to measure is $\cos\left(\alpha_{\text{opt}}\right)\sigma_{x}+\sin\left(\alpha_{\text{opt}}\right)\sigma_{y},$
where\\
$\alpha_{\text{opt}}\approx N\arctan\left[\tan\left(2g\tau\right)\cos\left(\theta_{\text{opt}}\right)\right]+\frac{\pi}{2}.$
This is an approximation as we neglect the the contribution of the classical (radial) part of the QFI. 
The dependence on the measurement basis is illustrated in fig. \ref{bases}.

\section{Spatially dependent coupling}\label{Spatially_dependent}
The following extends the results of the previous section to a more realistic scenario. We focus our analysis on the NV center, but it could be easily extended to other quantum sensors.
In reality, the simplified model interaction Hamiltonian \eqref{const2} should be replaced with the corresponding dipole-dipole interaction term,
\beq\label{spatial1}
H_1=\sigma_z\sum_{j=1}^Ng^j\left(\bar{r}_j\right)I_x^j.
\eeq
In spherical coordinate system, where the NV is situated at the origin and the NV's magnetization axis coincides with the $\hat{z}$ axis, denoting by $\left\{r_j,\theta_j,\varphi_j\right\}$ the coordinates of the j-th nucleus, which are also time dependent, the coupling constants can be written as,
\beq\label{spatial2}
g^j=-3 J r_j^{-3}\sin\theta_j\cos\theta_j\cos\varphi_j,
\eeq
with the physical coupling constant $J=\frac{\mu_0\hbar\gamma_e\gamma_N}{4\pi}=0.49 \textrm{MHz}\cdot\textrm{nm}^3$ , where $\mu_0$ is the vacuum permeability, $\hbar$ is the reduced plank constant and $\gamma_{e/N}$ are the electronic/nuclear gyro-magnetic ratios. The coupling constant can be rewritten in terms of the definitions in \ref{Appendix: DD} as
\beq\label{spatial3}
g^j=-J r_j^{-3}\left(\tilde{\zeta}_1 Y_2^{\left(1\right)}\left(\Omega_j\right)+\tilde{\zeta}_{-1}Y_2^{\left(-1\right)}\left(\Omega_j\right)\right),
\eeq
where $\Omega$ is the solid angle.
	 The protocol is identical, and the analysis follows the same lines as in the simplified model. Nevertheless, when writing the reduced density matrix of the sensor, eq. \eqref{constant} changes due to the spatial dependency of \eqref{spatial1} to
	 \beq\label{spatial4}
	  \langle\psi_{1}|\psi_{2}\rangle=\prod_{j=1}^{N}\left(\cos^2\left(\frac{\omega_N}{2}t\right)e^{2iG_j}+\sin^2\left(\frac{\omega_N}{2}t\right)e^{-2iG_j}\right),
	 \eeq
	 where $G_i=\int\limits_0^\tau dt g^j\left(t\right)$. We define the single nucleus module and phase as
	 \begin{align}\label{spatial5}
	  &\Phi_j=\arctan\left[\tan\left(2 G_j\right)\cos\left(\omega_N t\right)\right]\ \textrm{and}\\\label{spatial6}
	 &r_j=\sqrt{\cos^2\left(2G_j\right)+\sin^2\left(2G_j\right)\cos^2\left(\omega_N t\right)}=\sqrt{1-\sin^2\left(2G_j\right)\sin^2\left(\omega_N t\right)}.
 \end{align}
 Substituting \eqref{spatial5} and \eqref{spatial6} into \eqref{spatial4}, we end up with
 \beq\label{spatial7}
 \langle\psi_{1}|\psi_{2}\rangle=\prod_{j=1}^{N}\left(1-\sin^2\left(2G_j\right)\sin^2\left(\omega_N t\right)\right)^{1/2}\exp\left\{i \arctan\left[\tan\left(2 G_j\right)\cos\left(\omega_N t\right)\right]\right\}.
 \eeq
 Under the weak coupling assumption $G_j\ll1$ we can approximate \eqref{spatial7},
 \beq\label{spatial8}
 \prod_{j=1}^{N}\exp\left(-2G_j^2\sin^2\left(\omega_N t\right)\right)\exp\left\{i 2 G_j\cos\left(\omega_N t\right)\right\}=\exp\left(-2\sum_{j=1}^NG_j^2\sin^2\left(\omega_N t\right)\right)\exp\left(i 2\sum_{j=1}^N G_j\cos\left(\omega_N t\right)\right).
 \eeq
 When $N\gg1$,
 \beq\label{spatial9}
 \sum_{j=1}^N G_i\rightarrow \left<\sum_{j=1}^NG_j\right>=\int\limits_0^\tau dt\left<\sum_{j=1}^Ng^j\left(t\right)\right>=N\int\limits_0^\tau dt\left(\frac{1}{V}\int\limits d^3 r g\left(\bar{r}\right)\right)=n\tau \int d^3r g\left(\bar{r}\right)\equiv\gamma_{e}\left<B\right>\tau=-\pi n J \tau \sin\left(2\alpha\right),
 \eeq
 where we denoted by $n=\frac{N}{V}$ the nuclei number density, $\alpha$ is the NV's tilting angle and $g\left(\bar{r}\right)=J r^{-3}\left(\tilde{\zeta}_1 Y_2^{\left(1\right)}\left(\Omega\right)+\tilde{\zeta}_{-1}Y_2^{\left(-1\right)}\left(\Omega\right)\right)$. In the second equality of \eqref{spatial9} we assumed that the average magnetic field is time independent. This is equivalent to the assumption that the liquid is an incompressible fluid, since fluctuations in the average magnetic field of a fully polarized sample can only arise from density fluctuations. The average magnetic field \eqref{spatial9} is calculated in \ref{Mean_field}. In the last equality we cite the result \eqref{mean_field_planar},
  
 We evaluate the second moment by adding an average over realizations
 \beq
 \left<\sum_{j=1}^N G_j^2\right>\rightarrow N\int\limits_0^\tau dt'\int\limits_0^\tau dt''\left<g\left(\bar{r}\left(t'\right)\right)g\left(\bar{r}\left(t''\right)\right)\right>.\eeq
 Assuming that the correlation $C_2\left(t',t''\right)\equiv\left<g\left(\bar{r}\left(t'\right)\right)g\left(\bar{r}\left(t''\right)\right)\right>$ is stationary,
 \beq\label{spatial11}
   \left<\sum_{j=1}^N G_j^2\right>\rightarrow N\int\limits_0^\tau d\tilde{t}'\int\limits_{-\tilde{t}'}^{\tilde{t}'} d\tilde{t}''C_2\left(t''\right).
   \eeq
   The physical quantity associated with eq. \eqref{spatial11} is 
   \beq\label{spatial12}
   B_{rms}^2=\frac{N}{\gamma_e^2}\lim_{t\rightarrow0}C_2\left(t\right)=f_2\frac{n}{d^3}\left(\frac{\mu_0\gamma_N\hbar}{4\pi}\right)^2.
   \eeq
  The proportion factor $f_2$ in \eqref{spatial12} is a function of the NV's tilting angle and is equal to
  \beq\label{spatial13}
  f_2= \frac{35 \pi -3 \pi  \cos (4 \alpha )}{256}.
  \eeq
  The full calculation of \eqref{spatial13} is found in \ref{Decay_Planar}.
  The result of \eqref{spatial11} for an arbitrary time is quite complex \cite{microfludics}, but for times $\tau\ll\tau_D$, where $\tau_D=\frac{d^2}{D}$ is the diffusion time, it can be simplified to the instantaneous limit
  \beq\label{spatial14}
  \left<\sum_{j=1}^N G_j^2\right>\rightarrow\gamma_e^2B_{rms}^2\tau^2,
  \eeq
  while for long times $\tau\gg\tau_D$,
  \beq\label{spatial15}
   \left<\sum_{j=1}^N G_j^2\right>\rightarrow\gamma_e^2B_{rms}^2\tau\tau_D=\tau S\left(\omega=0\right),
  \eeq
  where $S\left(\omega\right)=\int\limits_{-\infty}^{\infty}dt C_2\left(t\right)e^{-i\omega t}$ is the power spectrum. 
  Assuming $\left|\left<B\right>\right|\gg B_{rms}$, which means that the net polarization is larger than the statistical one,  we can stop our expansion in moments of $G_j$ at second order. Since $B_{rms}\propto d^{-3/2}$, for an increasingly shallow NV, eventually $B_{rms}\sim\left|\left<B\right>\right|$, and than higher moment should be taken into account. Some calculations of the third moment can be found in \ref{Third_moment}.
  Using the definitions \eqref{spatial9} and \eqref{spatial12}, eq. \eqref{spatial8} can be written as
  \beq\label{spatial16}
   \langle\psi_{1}|\psi_{2}\rangle=\exp\left(-2\gamma_e^2 B_{rms}^2\tau^2\sin^2\left(\omega_N t\right)\right)\exp\left(i 2\gamma_e\left<B\right>\tau\cos\left(\omega_N t\right)\right),
  \eeq
  for times $\tau\ll\tau_D$. From eq. \eqref{spatial16} we identify the total accumulated phase $\Phi$ and decay $r$ as 
  \begin{align}\label{spatial17}
  &\Phi=2\gamma_e\left<B\right>\tau\cos\left(\omega_N t\right),\\\label{spatial18}
  &r\underbrace{\approx}_{\tau\ll\tau_D}\exp\left(-2\gamma_e^2 B_{rms}^2\tau^2\sin^2\left(\omega_N t\right)\right).
  \end{align}
  The QFI can now be estimated by substituting eqs. \eqref{spatial17} and \eqref{spatial18} into the formula \eqref{Bures},
  \beq\label{spatial19}
  \pazocal{I}=4 \gamma_e^2\tau^2t^2 \sin ^2(\theta ) e^{-4 \gamma _e^2 B_{rms}^2 \tau ^2  \sin ^2(\theta )}\left[\left<B\right>^2+B_{rms}^2\frac{4  \gamma _e^2 B_{rms}^2 \tau ^2   \cos ^2(\theta)}{1-e^{-4 \gamma _e^2 B_{\text{rms}}^2  \tau ^2\sin ^2(\theta )}}\right].
  \eeq
  The first term in \eqref{spatial19} is the information that can be extracted by the signal (or phase accumulation), while the second term is the information that comes from the noise. The latter's contribution is smaller by a factor of $\left(\frac{B_{rms}}{\left<B\right>}\right)^2$ and, therefore, negligible under the assumption $\left|\left<B\right>\right|\gg B_{rms}$.
  When the back-action is negligible, meaning $\gamma_e^2B_{rms}^2\tau^2\ll1$, the QFI is approximately
  \beq\label{spatial20}
  \pazocal{I}\approx4\gamma_e^2\tau^2t^2\sin^2\left(\theta\right)\left<B\right>^2.
  \eeq
   Hence, the QFI \eqref{spatial20} is optimal for $\sin^2\theta=1$, for which
   \beq\label{spatial21}
    \pazocal{I}\approx4\gamma_e^2\tau^2t^2\left<B\right>^2.
   \eeq 
   Eq. \eqref{spatial21} is equivalent to the eq. \eqref{toy_optimal_weak} in sec. \ref{Simplified}, with $N g \tau$ replaced by the $\gamma_e \left<B\right>\tau$. When the back-action is strong, the optimal $\theta$ changes such that,
  \beq\label{optimal_cond}
  \sin^2\left(\theta\right)=\frac{1}{4\gamma_e^2B_{rms}^2\tau^2},
  \eeq
  then the QFI is
  \beq\label{spatial22}
  \pazocal{I}\approx\frac{\left<B\right>^2}{e B_{rms}^2}t^2\approx 17.5\frac{nd^3}{e}t^2,
  \eeq
  where in the last equality we substituted an NV's tilting angle of $\alpha=54.7^\circ$.
    This is the same result as in the previous section, with the difference that due to the dipolar interaction and the geometry, there is a dependence on the NV's tilting angle and the effective number of particles is $nd^3$.
   When $\tau\gg\tau_D$ the decay changes to
  \beq\label{spatial23}
  r\underbrace{\approx}_{\tau\gg\tau_D}e^{-2\gamma_e^2B_{rms}^2\tau\tau_D\sin^2\left(\omega_Nt\right)}.
  \eeq
  The QFI given \eqref{spatial17} and \eqref{spatial23} is
  \beq\label{spatial24}
 \pazocal{I}=4 \gamma_e^2\tau^2t^2 \sin ^2(\theta ) e^{-4 \gamma _e^2 B_{rms}^2 \tau \tau_D  \sin ^2(\theta )}\left[\left<B\right>^2+B_{rms}^2\frac{4  \gamma _e^2 B_{rms}^2 \tau_D ^2   \cos ^2(\theta)}{1-e^{-4 \gamma _e^2 B_{\text{rms}}^2  \tau \tau_D \sin ^2(\theta )}}\right].
  \eeq
  Still under the assumption $\left|\left<B\right>\right|\gg B_{rms}$ the noise contribution can be neglected. Since the signal \eqref{spatial17} doesn't change, as the average field does not depend on diffusion, the QFI also remains the same in the small back-action limit $\gamma_e^2B_{rms}^2\tau\tau_D\ll1$, where the decay is negligible. Hence the small back-action limit recovers \eqref{spatial21}.
  Strong back-action will lead to a change in \eqref{optimal_cond}, since the decay dictates the optimal $\theta$, 
  \beq\label{optimal_cond2}
  \sin^2\left(\theta\right)=\frac{1}{4\gamma_e^2 B_{rms}^2\tau\tau_D}.
  \eeq 
  Substituting \eqref{optimal_cond2} into \eqref{spatial24} yields the optimal QFI, 
  \beq\label{long_times_opt}
  \pazocal{I}=\frac{\left<B\right>^2}{e B_{rms}^2}\frac{\tau}{\tau_D}t^2.
  \eeq
  We note that eq. \eqref{long_times_opt} is problematic since the QFI grows linearly with $\tau$ indefinitely. More over, if $D\rightarrow\infty$ the QFI diverges. The latter is explained by the fact that for $\tau_D\rightarrow0$ the back-action is always small, therefore, the relevant QFI is \eqref{spatial21} and not \eqref{long_times_opt}. The first problem arises from the fact that we took into account an infinite volume with an infinite number of spins. Hence, as more spins pass through the effective interaction region the information increases. We have to take into account finite volume effects when the QFI \eqref{long_times_opt} is no longer in accord with the standard quantum limit, namely,
  \beq\label{spatial25}
  \frac{\left<B\right>^2}{e B_{rms}^2}\frac{\tau}{\tau_D}t^2>Nt^2.
  \eeq
  Eq. \eqref{spatial25} leads to the following condition on the interrogation time $\tau$,
  \beq\label{spatial26}
  \tau\geq \frac{1}{17.5}\tau_D \left(\frac{V}{d^3}\right).
  \eeq
  Perhaps, a more intuitive form for \eqref{spatial26} can be written in terms of the volumetric diffusion time $\tau_V\equiv \frac{V^{2/3}}{D}$,
  \beq\label{spatial27}
  \tau\geq\frac{1}{17.5}\tau_V \left(\frac{V^{1/3}}{d}\right),
  \eeq 
  and since usually $\frac{V^{1/3}}{17.5d}>1$ \eqref{spatial27} can be regarded as $\tau\gg\tau_V$. The volumetric diffusion time is the approximate time it takes for a nucleus to travel from one end of the volume to the other by diffusion. Therefore, for times bigger than this, it is clear that boundary effect will start to play an important role. For such long times the diffusion propagator approaches to uniform distribution over the volume, which changes the decay to 
  \beq\label{spatial28}
  r\underbrace{\approx}_{\tau\gg\tau_V}e^{-\frac{2}{nV}\gamma_e^2\left<B\right>^2\tau^2}.
  \eeq  
  The result \eqref{spatial28} is a general property, as the propagator should always approach the uniform distribution for times $\tau\gg\tau_V$.
  The QFI can again be calculated using the formula \eqref{Bures} with \eqref{spatial17} and \eqref{spatial28},
  \beq
  \pazocal{I}=4 t^2 \gamma _e^2 \left<B\right>^2 \tau ^2 \sin ^2(\theta ) e^{-\frac{4 \gamma _e^2 \left<B\right>^2 \tau ^2  \sin ^2(\theta )}{N}}\left[1+\frac{4 \gamma _e^2 \left<B\right>^2  \tau ^2 \cos ^2\left(\theta \right) }{N^2 \left(1-e^{-\frac{4 \gamma _e^2 \left<B\right>^2 \tau ^2 \sin ^2(\theta )}{N}}\right)}\right].
  \eeq
  The noise term is negligible as even for strong back-action $\gamma_e^2\left<B\right>^2\tau^2/N\gg1$ it is decreased by a factor of $1/N$. The optimal QFI is achieved in the strong back-action limit for
  \beq\label{optimal_cond3}
  \sin\left(\phi\right)=\frac{N}{4\gamma_e^2\left<B\right>^2\tau^2},
  \eeq
  for which the QFI is
  \beq\label{spatial29}
  \pazocal{I}\approx\frac{N}{e}t^2.
  \eeq
  Eq. \eqref{spatial29} is equivalent to eq. \eqref{toy_optimal_strong}, meaning that  effectively at long interrogation time, the dipolar interaction acts as a constant coupling $g=\frac{\gamma_e \left<B\right>}{N}$.

  \section{Undriven nuclei}\label{undriven}
\begin{figure}[t]
	\begin{center}
		\includegraphics[width=0.48\textwidth]{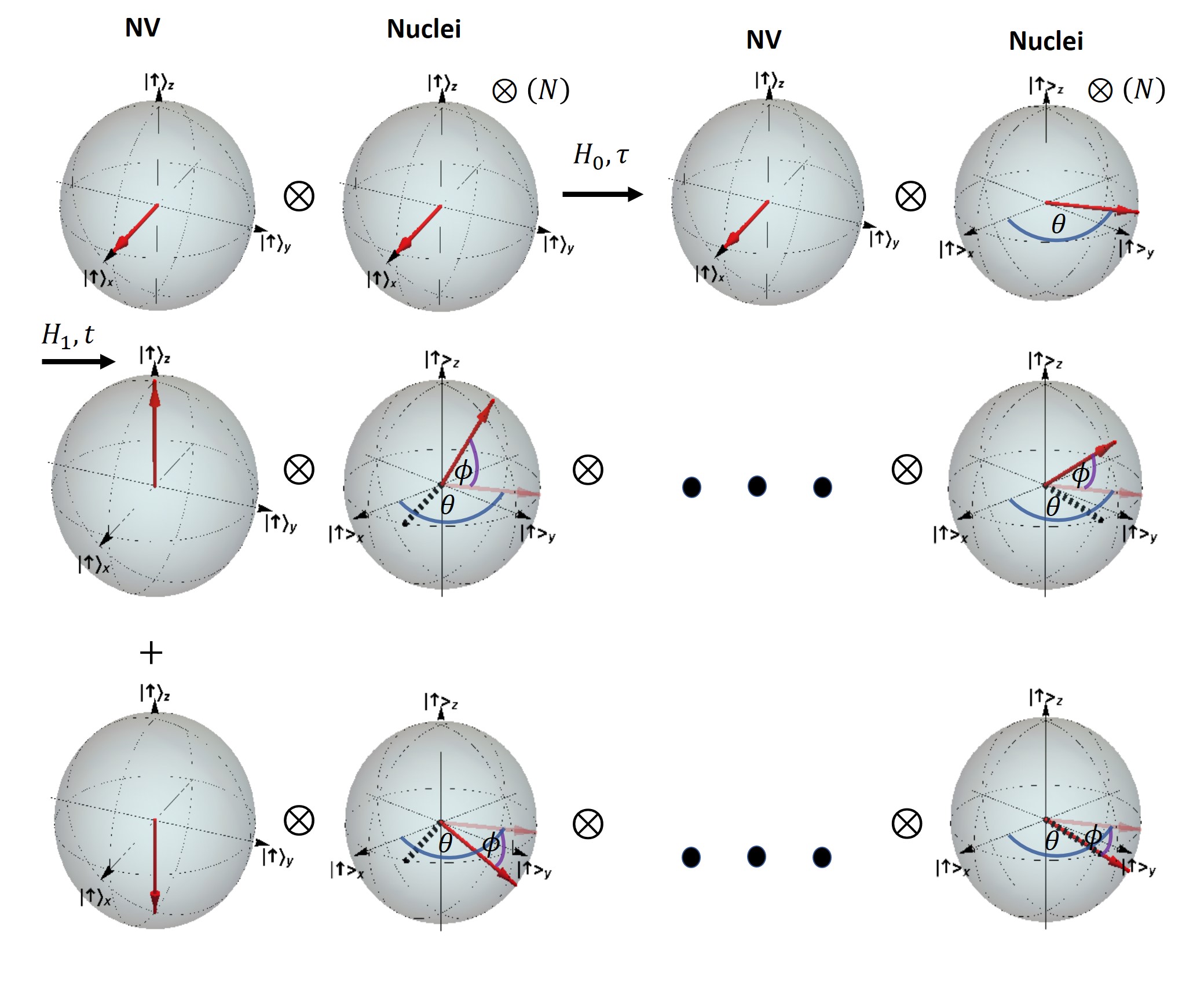}
	\end{center}
	\caption{ The NV and the nuclear spins are initialized at the $x$ direction. The nuclei are then allowed, using an appropriate pulse sequence, to propagate according to the free Hamiltonian $H_0$ for a time $t$, which results in a rotation by $\theta=\omega_N t$ around the $z$ axis. Then, by changing the external pulse sequence, the system propagates under the interaction Hamiltonian $H_1$ for a duration of $\tau$. This results in a rotation of each nucleus around a random axis in the $x-y$ plane (dashed line) by an angle that depends on the NV's state and the rotation axis. The randomness of the rotation axis will generate entanglement between the nuclei and the NV which will cause an effective dephasing of the NV. We refer to this effect as "classical| dephasing" as it arises from the classical uncertainty of the rotation axis.}
	\label{UD_dephasing}
\end{figure}

  In the following we would like to drop the assumption that we drive the nuclei in the $\hat{x}$ direction in order to achieve \eqref{spatial1}. This is an interesting scenario both experimentally, as applying an external drive on the nuclei might prove to be challenging, and analytically since it will introduce another type of dephasing to the quantum probe. 
  The free Hamiltonian of the system is
  \beq\label{nodrive1}
  H_0=\frac{\omega_0}{2} \sigma_z+\frac{\omega_N}{2} \sum\limits_{i=1}^{N} I_z,
  \eeq
  and the full dipolar interaction Hamiltonian is
  \beq\label{nodrive2}
  H_{DD}=\sum_{i,j\in\left\{x,y,z\right\}}\sigma_i\sum_{k=1}^N g_{i,j}^kI_j^k
  \eeq
	The interaction Hamiltonian \eqref{nodrive2} in the interaction picture with respect to the NV's energy gap is
  \beq\label{nodrive3}
  H_1\approx \sigma_z\sum_{i=1}^N\left(g_0^i I_z^i+g_1^i  I_+^i +g_2^i  I_-^i\right),
  \eeq
  where we made the rotating wave approximation $g_{i,j}^k\ll\omega_0$.
  As customary in spectrum measurements, the NV is driven with $\pi$-pulses every time $\tau_p$, so that the effective Hamiltonian \eqref{nodrive3} is
  \beq\label{nodrive4}
    H_1=h(t)\sigma_z\left(g_0^i I_z^i+g_1^i  I_+^i +g_2^i  I_-^i\right),
  \eeq
  where
  \beq\label{nodrive5}
  h(t)=\frac{1}{(N+1)\tau_p}\sum_{n=0}^N \left(-1\right)^n\Theta\left(t-n\tau_p\right)\Theta\left(t-(n+1)\tau_p\right)
  \eeq
  We expand the function $h$ into a Fourier series, 
  \beq\label{nodrive6}
  h(t)=\sum_{m=-\infty}^\infty a_m e^{im\omega t}
  \eeq
  Substituting the expansion \eqref{nodrive6} into \eqref{nodrive5} gives an equation for the coefficients $a_m$, 
  \beq\label{nodrive7}
  \sum_{m=-\infty}^\infty a_m e^{im\omega t}=\frac{1}{(N+1)\tau_p}\sum_{n=0}^N \left(-1\right)^n\Theta\left(t-n\tau_p\right)\Theta\left(t-(n+1)\tau_p\right).
  \eeq
  Multiplying both sides of \eqref{nodrive7} by $e^{-i k \omega t}$ and integrating,
  \begin{align}\label{nodrive8}
  a_k&=\sum_{n=0}^N \left(-1\right)^n\frac{1}{(N+1)\tau_p}\int\limits_{n\tau_p}^{(n+1)\tau_p}dt e^{-ik\omega t}=\frac{i}{k(N+1)\omega\tau_p}\left(e^{-ik\omega \tau_p}-1\right)\sum_{n=0}^N \left(-1\right)^ne^{-ik n\omega \tau_p},
  \end{align}
  where in the last equality we assumed $k\in\mathbb{Z}/\left\{0\right\}$. 
  We set $\omega=\omega_p\equiv\frac{\pi}{\tau_p}$, the pulse frequency, for which the expression simplifies and is non-zero only for odd k,
 \begin{align}\label{nodrive9}
 a_k&=\frac{i}{\pi k(N+1)}\left(\left(-1\right)^k-1\right)\sum_{n=0}^N \left(-1\right)^{n\left(k+1\right)}\underbrace{=}_{\textrm{k\ odd}}-\frac{2i}{k\pi(N+1)}\sum_{n=0}^N 1=-\frac{2i}{\pi k}.
 \end{align}
 To determine $a_0$, we integrate both sides of eq. \eqref{nodrive7},
 \begin{align}\label{nodrive10}
 a_0&=\sum_{n=0}^N\left(-1\right)^n\frac{1}{(N+1)\tau_p}\int\limits_{n\tau_p}^{(n+1)\tau_p}dt 1=\frac{1}{N+1}\sum_{n=0}^N(-1)^n=\begin{cases}
 0 & \textrm{N\ odd}\\
 1 & \textrm{N\ even}
 \end{cases}.
 \end{align}
	We continue under the assumption that N is odd. In this case, substituting the results \eqref{nodrive9} and \eqref{nodrive10} into the expansion \eqref{nodrive6},
	\beq\label{nodrive11}
	h\left(t\right)=-\frac{2i}{\pi}\sum_{n=-\infty}^\infty\frac{1}{2n+1}e^{i(2n+1)\omega_pt}.
	\eeq
	The Hamiltonian \eqref{nodrive4}, can now be rewritten using \eqref{nodrive11} in the interaction picture with respect to the nuclear spins energy gaps,
  \beq\label{nodrive12}
  H_1=-\frac{2i}{\pi} \sum_{n=-\infty}^{\infty}\frac{1}{2n+1}e^{i(2n+1)\omega_p t}\left[g_0^i \sigma_z I_z^i+g_1^i \sigma_z I_+^ie^{i\omega_N t} +g_2^i \sigma_z I_-^ie^{-i\omega_N t}\right],
  \eeq
  When $\omega\sim\omega_N$ such that $\delta\omega\equiv\omega_N-\omega\ll\omega_N$ we can approximate \eqref{nodrive12}, assuming  $\omega_N\gg g_0^i,g_1^i,g_2^i$,
  \beq\label{nodrive13}
  H_1\approx-\frac{2i}{\pi} \sigma_z\left[-g_1^i  I_+^ie^{i\delta\omega t} +g_2^i  I_-^ie^{-i\delta\omega t}\right]=-\frac{2g^i_{\pm}}{\pi} \sigma_z\left[  I_x^i\sin\left(\delta\omega t-\varphi_i\right) + I_y^i\cos\left(\delta\omega t-\varphi_i\right)\right],
  \eeq
  where we used the explicit spatial forms $g_k^j=-\frac{3}{2}r_j^{-3}\cos\theta_j\sin\theta_je^{\left(-1\right)^ki\varphi_j}$ (see sec. \ref{Appendix: DD}) and denoted $g_{\pm}^j= -\frac{3}{2}r_j^{-3}\cos\theta_j\sin\theta_j$.
  We now follow our previous protocol using the effective interaction \eqref{nodrive13}. The system is initialized to the state
  \beq\label{nodrive14}
  \ket{\psi_i}=\ket{\uparrow_X}^S\otimes\ket{\uparrow_X\cdots\uparrow_X}.
  \eeq
  We shall work in the interaction with respect to $H_{ref}=\frac{\omega_0}{2}\sigma_z+\frac{\omega_p}{2}\sum_{j=1}^NI_z^j$, in which the free Hamiltonian \eqref{nodrive1} and the effective interaction Hamiltonian \eqref{nodrive13}  are
  \begin{align}\label{nodrive15}
 & H_0=\frac{\delta\omega}{2}\sum_{i=1}^N I^i_z \\\label{nodrive16}
 &H_1=\frac{2g^i_{\pm}}{\pi} \sigma_z\sum_{i=1}^N\left[ I_x^i\sin\left(\varphi_i\right) - I_y^i\cos\left(\varphi_i\right)\right]\equiv\frac{2g^i_{\pm}}{\pi} \sigma_z\sum_{i=1}^Nh_i.
  \end{align}
  The system \eqref{nodrive14} is allowed to evolve for a time $t$ under the Hamiltonian $H_0$, \eqref{nodrive15}, 
  \beq\label{nodrive19}
  \ket{\psi_t}=\ket{\uparrow_X}^S\left(\cos\left(\frac{\delta\omega}{2} t\right)\ket{\uparrow_X}-i\sin\left(\frac{\delta\omega}{2} t\right)\ket{\downarrow_X}\right)^{\otimes N}
  \eeq 
  The eigenvectors of the single nucleus  Hamiltonian, $h_i=I_x^i\sin\left(\varphi_i\right)-I_y^i\cos\left(\varphi_i\right)$, are
  \begin{align}\label{nodrive17}
  	\ket{\uparrow_{\varphi_i}}&=\frac{1}{\sqrt{2}}\left(e^{-\frac{i}{2}\left(\varphi_i+\pi/2\right)}\ket{\uparrow_Z}-e^{\frac{i}{2}\left(\varphi_i+\pi/2\right)}\ket{\downarrow_Z}\right),\ \textrm{with corresponding eigenvalue}\ 1\\
  	\ket{\downarrow_{\varphi_i}}&=\frac{1}{\sqrt{2}}\left(e^{-\frac{i}{2}\left(\varphi_i+\pi/2\right)}\ket{\uparrow_Z}+e^{-\frac{i}{2}\left(\varphi_i+\pi/2\right)}\ket{\downarrow_Z}\right),\ \textrm{with corresponding eigenvalue}\ -1 .
  \end{align}
Eq. \eqref{nodrive17}  can be rewritten in the $x$ basis,
  \begin{align}\label{nodrive18}
  \ket{\uparrow_{\varphi_i}}&=-i\sin\left(\frac{\varphi_i}{2}+\frac{\pi}{4}\right)\ket{\uparrow_X}+\cos\left(\frac{\varphi_i}{2}+\frac{\pi}{4}\right)\ket{\downarrow_X},\ \textrm{with}\ \lambda=1\\
  \ket{\downarrow_{\varphi_i}}&=\cos\left(\frac{\varphi_i}{2}+\frac{\pi}{4}\right)\ket{\uparrow_X}-i\sin\left(\frac{\varphi_i}{2}+\frac{\pi}{4}\right)\ket{\downarrow_X},\ \textrm{with}\ \lambda=-1 .
  \end{align}
  The reverse transformation to \eqref{nodrive18} is
  \begin{align}\label{nodrive20}
  \ket{\uparrow_X}=\cos\left(\frac{\varphi_i}{2}+\frac{\pi}{4}\right)\ket{\downarrow_{\varphi_i}}+i\sin\left(\frac{\varphi_i}{2}+\frac{\pi}{4}\right)\ket{\uparrow_{\varphi_i}}\\
  \ket{\downarrow_X}=i\sin\left(\frac{\varphi_i}{2}+\frac{\pi}{4}\right)\ket{\downarrow_{\varphi_i}}+\cos\left(\frac{\varphi_i}{2}+\frac{\pi}{4}\right)\ket{\uparrow_{\varphi_i}}.
  \end{align}
  We can write \eqref{nodrive19} in the eigenbasis \eqref{nodrive18} using \eqref{nodrive20},
  \begin{equation}\label{nodrive21}
 \ket{\psi_t}=\ket{\uparrow_X}^S\left(\cos\left(\frac{\delta\omega}{2} t-\frac{\varphi_i}{2}-\frac{\pi}{4}\right)\ket{\downarrow_{\varphi_i}}-i\sin\left(\frac{\delta\omega}{2} t-\frac{\varphi_i}{2}-\frac{\pi}{4}\right)\ket{\uparrow_{\varphi_i}}\right)^{\otimes N}.
  \end{equation}
  Now we can easily propagate \eqref{nodrive21} according to the interaction Hamiltonian \eqref{nodrive16}  for a time period $\tau$,
  \begin{align}\label{nodrive22}
  \ket{\psi_\tau}=\frac{1}{\sqrt{2}}\ket{\uparrow_Z}^S\left(\cos\left(\frac{\delta\omega}{2} t-\frac{\varphi_i}{2}-\frac{\pi}{4}\right)e^{i \frac{2G_{\pm}^i}{\pi}}\ket{\downarrow_{\varphi_i}}-e^{-i \frac{2G_{\pm}^i}{\pi}}i\sin\left(\frac{\delta\omega}{2} t-\frac{\varphi_i}{2}-\frac{\pi}{4}\right)\ket{\uparrow_{\varphi_i}}\right)^{\otimes N}\\\nonumber
  +\frac{1}{\sqrt{2}}\ket{\downarrow_Z}^S\left(\cos\left(\frac{\delta\omega}{2} t-\frac{\varphi_i}{2}-\frac{\pi}{4}\right)e^{-i \frac{2G_{\pm}^i}{\pi}}\ket{\downarrow_{\varphi_i}}-e^{i \frac{2G_{\pm}^i}{\pi}}i\sin\left(\frac{\delta\omega}{2} t-\frac{\varphi_i}{2}-\frac{\pi}{4}\right)\ket{\uparrow_{\varphi_i}}\right)^{\otimes N},
  \end{align}
  Where $G_{\pm}^i=\int\limits_0^\tau dt g_\pm^i\left(t\right)$.
Eq. \eqref{nodrive22} leads us to define the states $\ker{\psi_1}$ and $\ket{\psi_2}$ by   $\ket{\psi_\tau}\equiv\frac{1}{\sqrt{2}}\left(\ket{\uparrow_Z}^S\ket{\psi_1}+\ket{\downarrow_Z}^S\ket{\psi_2}\right)$. Using \eqref{nodrive22} we can calculate the overlap, 
 \begin{align}\label{nodrive23}
 \left<\psi_2|\psi_1\right>&=\prod_{i=1}^N\left[\cos^2\left(\frac{\delta\omega}{2} t-\frac{\varphi_i}{2}-\frac{\pi}{4}\right)e^{i\frac{4G_{\pm}^i}{\pi}}+\sin^2\left(\frac{\delta\omega}{2} t-\frac{\varphi_i}{2}-\frac{\pi}{4}\right)e^{-i\frac{4G_{\pm}^i}{\pi}}\right]\\\nonumber
 &=\prod_{i=1}^N\left[\cos\left(\frac{4G_{\pm}^i}{\pi}\right)+i\sin\left(\frac{4G_{\pm}^i}{\pi}\right)\sin\left(\delta\omega t-\varphi_i\right)\right].
 \end{align}
 We can write each term in the product \eqref{nodrive23} in a polar representation
 \begin{align}\label{nodrive24}
 \Phi_i&=\arctan\left[\tan\left(\frac{4G_{\pm}^i}{\pi}\right)\sin\left(\delta\omega t-\varphi_i\right)\right],\\\label{nodrive25}
 r_i&=\sqrt{1-\sin^2\left(\frac{4G_{\pm}^i}{\pi}\right)\cos^2\left(\delta\omega t-\varphi_i\right)}.
 \end{align}
In the weak coupling regime, $g_{\pm}^i\tau\ll1$, eqs. \eqref{nodrive24} and \eqref{nodrive25} can be approximated,
\begin{align}\label{nodrive26}
\Phi_i&\approx\frac{4G_{\pm}^i}{\pi}\sin\left(\delta\omega t-\varphi_i\right),\\\label{nodrive27}
r_i&=\exp\left(-8\left(\frac{G_{\pm}^i}{\pi}\right)^2\cos^2\left(\delta\omega t-\varphi_i\right)\right).
\end{align}
The total signal $\Phi=\sum_{i=1}^N\Phi_i$, can be evaluated by the average as $N\rightarrow\infty$,
\begin{align}
\Phi\approx\left<\sum_{i=1}^N\Phi_i\right>&=\frac{4 n \tau}{\pi}J\int \frac{d^3r}{r^3} (-\frac{3}{2})\cos\theta\sin\theta\sin\left(\delta\omega t-\varphi\right)\\\nonumber
&=\frac{4 n \tau}{\pi}J\left[-\frac{3}{4i}e^{i\delta\omega t}\int \frac{d^3r}{r^3} \cos\theta\sin\theta e^{-i\varphi}+\frac{3}{4i}e^{-i\delta\omega t}\int \frac{d^3r}{r^3} \cos\theta\sin\theta e^{i\varphi}\right]\\\nonumber
&=\frac{4 n \tau}{\pi}J\frac{1}{2i}\left[\tilde{\zeta}_{-1}e^{i\delta\omega t}\int \frac{d^3r}{r^3} Y_2^{\left(-1\right)}\left(\Omega\right)-\tilde{\zeta}_{1}e^{-i\delta\omega t}\int \frac{d^3r}{r^3}Y_2^{\left(1\right)}\left(\Omega\right)\right],
\end{align}
where in the last equality we used the notations in \eqref{SH_coeff}. 
We can identify these integrals as $I_1^{\pm1}$ from sec. \ref{Sec_mean_planar}, 
\begin{align}\label{nodrive28}
\Phi\approx\frac{4 n \tau}{\pi}J I_1^{\left(1\right)}\sin\left(\delta\omega t\right)
=-2 n \tau J\sin\left(\delta\omega t\right)\sin\left(2\alpha\right)\equiv\gamma_e\left<B\right>\tau\sin\left(\delta\omega t\right). 
\end{align}
The mean field \eqref{nodrive28} is the same as \eqref{spatial9} up to a prefactor of $\frac{2}{\pi}$. 
The total decay can be evaluated at times $\tau\ll\tau_D$ by the instantaneous correlation function,
\begin{align}\label{nodrive29}
&C_2^0\left(t\right)\equiv C_2\left(\tau=0,t\right)=\frac{8 n}{\pi^2}\left(\frac{\hbar\mu_0 \gamma_N}{4\pi}\right)^2\left(-\frac{3}{2}\right)^2\int\frac{d^3r}{r^6}\cos^2\theta\sin^2\theta\cos^2\left(\delta\omega t-\varphi\right)\\\nonumber
&=\frac{8 n}{\pi^2}\left(\frac{\hbar\mu_0 \gamma_N}{4\pi}\right)^2\left(-\frac{3}{4}\right)^2\left[e^{2i \delta\omega t}\int\frac{d^3r}{r^6}\cos^2\theta\sin^2\theta e^{-2i\varphi}+e^{-2i \delta\omega t}\int\frac{d^3r}{r^6}\cos^2\theta\sin^2\theta e^{2i\varphi}+2\int\frac{d^3r}{r^6}\cos^2\theta\sin^2\theta \right]\\\nonumber
&=\frac{4 n}{\pi^2}\left(\frac{\hbar\mu_0 \gamma_N}{4\pi}\right)^2\left[\cos\left(2\delta\omega t\right) I_2^{\left(1,1\right)}+I_2^{\left(1,-1\right)}\right],
\end{align}
where the integrals (see sec. \ref{Decay_Planar})
\begin{align}\label{Int1}
I_2^{\left(1,1\right)}\left(\alpha\right)&=\frac{-28 \pi  \cos (2 \alpha )+13 \pi  \cos (4 \alpha )+15 \pi }{2048 d^3},\\\label{Int2}
I_2^{\left(1,-1\right)}\left(\alpha\right)&=\frac{28 \pi  \cos (2 \alpha )+13 \pi  \cos (4 \alpha )+87 \pi }{2048 d^3}.
\end{align}
We can rewrite \eqref{nodrive29} so it will be more similar to the previous section,
\begin{align}\label{nodrive31}
C_2^0\left(t\right)&=\frac{4 n}{\pi^2}\left(\frac{\hbar\mu_0  \gamma_N}{4\pi}\right)^2\left[2\cos^2\left(\delta\omega t\right) I_2^{\left(1,1\right)}+I_2^{\left(1,-1\right)}-I_2^{\left(1,1\right)}\right].
\end{align}
The decay consists of two terms - the oscillating part that is associated with entanglement and the constant decay due to classical dephasing. The idea of classical dephasing is demonstrated in fig. \ref{UD_dephasing} - the interaction Hamiltonian causes each nucleus to rotate around a random axis in the $xy$ plane. The randomness inflicts an effective dephasing on the NV. This dephasing is mitigated by the NV's tilting, which dictates a preferred direction and effectively restricts the rotation axis only to a section of the Bloch sphere. This can be seen directly from eqs. \eqref{Int1}, \eqref{Int2} and \eqref{nodrive31}, where the oscillating part of the decay vanishes for $\alpha=0$ while the classical dephasing is maximal. 
We note that $C_2^0\left(0\right)=\frac{2}{\pi^2}\left(B_{rms}^2\right)_{with\ drive}$.
The decay for times $\tau\ll\tau_D$ is therefore,
\beq
r\approx e^{-\gamma_e^2C_2^0\left(t\right)\tau^2}.
\eeq
If the back-action $\gamma_eB_{rms}\tau\ll1$ then we retrieve \eqref{spatial21} with the mean field \eqref{nodrive28}. For strong back-action, there are two apparent choices. We can choose to follow our previous logic and set,
\beq\label{opt_cond_no_drive}
\cos^2\left(\delta\omega t\right)=\frac{\pi^2}{8n\tau^2 \left(\frac{\hbar\mu_0 \gamma_e \gamma_N}{4\pi}\right)^2\left|I_2^{(1,1)}\right|},
\eeq
which is the analog to eq. \eqref{optimal_cond}.
The QFI for the times $t$ that follow \eqref{opt_cond_no_drive} will be
\beq\label{nodrive30}
\pazocal{I}\approx \frac{1}{\left|I_2^{(1,1)}\right|e^{\textrm{sgn}\left(I_2^{(1,1)}\right)}} \left(\frac{\left<B\right>}{B_{rms}}\right)^2_{\textrm{with\ drive}} e^{-\gamma_e^2 \tau^2 C_2^0\left(t=\frac{\pi}{2\delta\omega}\right)},
\eeq
where $\textrm{sgn}\left(x\right)$ is the sign function and the values of the mean field and $B_{rms}$ in $\left(\frac{\left<B\right>}{B_{rms}}\right)^2_{\textrm{with\ drive}}$ are given by eqs. \eqref{spatial9} and \eqref{spatial12}. The QFI \eqref{nodrive30} has an exponential decrease compared to eq. \eqref{spatial22}, due to the classical dephasing, and a decrease by a factor of $\left|I_2^{(1,1)}\right|$, because less entanglement was generated. This former can be mitigated by choosing a sufficiently small $\tau$, like $\tau^2=\frac{1}{\gamma_e^2C_2^0\left(t=\pi/2\delta\omega\right)}$, for which the QFI's differ only by a factor of $e$. Another strategy could be to maximize the signal with $\cos^2\left(\delta\omega t\right)=1$ 
and choose 
\beq\label{nodrive32}
\tau=\frac{1}{\gamma_e\sqrt{C_2^0\left(t=0\right)}}=\frac{\pi}{\sqrt{2}\gamma_e\left(B_{rms}\right)_{with\ drive}},
\eeq
for which
\beq\label{nodrive33}
\pazocal{I}=\frac{2}{e}\left(\frac{\left<B\right>}{B_{rms}}\right)^2_{\textrm{with\ drive}}.
\eeq

A caveat should be made about the eqs. \eqref{nodrive32} and \eqref{nodrive33}, as it appears that even at the presence of classical dephasing, the same QFI scaling can be achieved. While this statement is theoretically true, eq. \eqref{nodrive32} suggests it might take a very long interrogation time, far greater than the NV's $T_2$ time. For water density, $n=33 \textrm{nm}^{-3}$, an NV depth of $d=30\textrm{nm}$ and tilting angle of $54.7^\circ$ eq. \eqref{nodrive31} leads to $\gamma_eB_{rms}=5.5\textrm{KHz}$, and therefore by \eqref{nodrive32} $\tau\approx 183 \mu\textrm{s}$.

We would like to conclude this section with a remark about the initial basis. If we repeat the derivation in the Y basis eqs. \eqref{nodrive28} and \eqref{nodrive31} change to
\begin{align}
&\Phi=2 n \tau\frac{\hbar\mu_0 \gamma_e \gamma_N}{4\pi}\cos\left(\delta\omega t\right)\sin\left(2\alpha\right)\equiv\gamma_e\left<B\right>\tau\sin\left(\delta\omega t\right),\\
&C_2^0\left(t\right)=\frac{4 n}{\pi^2}\left(\frac{\hbar\mu_0  \gamma_N}{4\pi}\right)^2\left[2\sin^2\left(\delta\omega t\right) I_2^{\left(1,1\right)}+I_2^{\left(1,-1\right)}-I_2^{\left(1,1\right)}\right].
\end{align}
which is are a bit more similar to the signal and decay with the external drive.

\section{Finite polarization}
The following extends the ideas of sec. \ref{Spatially_dependent} in order to consider the more general case, where the spin's polarization is finite. The initial density matrix of the system
\beq\label{pol1}
\rho_i=\ket{\uparrow_X}\bra{\uparrow_X}_{S}\left(p\ket{\uparrow_X}\bra{\uparrow_X}+\left(1-p\right)\ket{\downarrow_X}\bra{\downarrow_X}\right)^{\otimes N},
\eeq  
where we denoted the probability of a single nucleus to be found at the $\ket{\uparrow_X}$ state as $p$ and accordingly $0\leq p\leq1$. Repeating our protocol, we find the evolution of \eqref{pol1} under the free Hamiltonian \eqref{const1} for a time $t$,
\begin{align}\label{pol2}
\rho_t=\ket{\uparrow_X}\bra{\uparrow_X}_{S}&\left[p\left(\cos\left(\frac{\omega_N}{2}t\right)\ket{\uparrow_X}-i\sin\left(\frac{\omega_N}{2}t\right)\ket{\downarrow_X}\right)\left(\cos\left(\frac{\omega_N}{2}t\right)\bra{\uparrow_X}+i\sin\left(\frac{\omega_N}{2}t\right)\bra{\downarrow_X}\right)\right.\\\nonumber
&\left.+\left(1-p\right)\left(\cos\left(\frac{\omega_N}{2}t\right)\ket{\downarrow_X}-i\sin\left(\frac{\omega_N}{2}t\right)\ket{\uparrow_X}\right)\left(\cos\left(\frac{\omega_N}{2}t\right)\bra{\downarrow_X}+i\sin\left(\frac{\omega_N}{2}t\right)\bra{\uparrow_X}\right)\right]^{\otimes N}
\end{align}
\begin{align*}
=\frac{1}{2}\ket{\uparrow_X}\bra{\uparrow_X}_{S}&\left[\left(1+\left(2p-1\right)\cos\left(\omega_N t\right)\right)\ket{\uparrow_X}\bra{\uparrow_X}+\left(1-\left(2p-1\right)\cos\left(\omega_N t\right)\right)\ket{\downarrow_X}\bra{\downarrow_X}\right.\\\nonumber
&\left.+i\left(2p-1\right)\sin\left(\omega_N t\right)\ket{\uparrow_X}\bra{\downarrow_X}-i\left(2p-1\right)\sin\left(\omega_N t\right)\ket{\downarrow_X}\bra{\uparrow_X}\right]^{\otimes N}.
\end{align*}
We then propagate the system \eqref{pol2} according to the Hamiltonian \eqref{spatial1}, $H_1=\sigma_z\sum_{j=1}^N g^jI_x^j$, for a time $\tau$, denoting $G_j=\int\limits_{0}^{\tau}g^j(t')dt'$,
\begin{align}\label{pol3}
\rho_{\tau}=\frac{1}{4}\ket{\uparrow_Z}\bra{\uparrow_Z}_{S}&\prod_{j=1}^N\otimes\left[\left(1+\left(2p-1\right)\cos\left(\omega_N t\right)\right)\ket{\uparrow_X}\bra{\uparrow_X}+\left(1-\left(2p-1\right)\cos\left(\omega_N t\right)\right)\ket{\downarrow_X}\bra{\downarrow_X}\right.\\\nonumber
&\left.+i\left(2p-1\right)\sin\left(\omega_N t\right)e^{-2iG_j}\ket{\uparrow_X}\bra{\downarrow_X}-i\left(2p-1\right)\sin\left(\omega_N t\right)e^{2iG_j}\ket{\downarrow_X}\bra{\uparrow_X}\right]
\end{align}
\begin{align*}
+\frac{1}{4}\ket{\downarrow_Z}\bra{\downarrow_Z}_{S}&\prod_{j=1}^N\otimes\left[\left(1+\left(2p-1\right)\cos\left(\omega_N t\right)\right)\ket{\uparrow_X}\bra{\uparrow_X}+\left(1-\left(2p-1\right)\cos\left(\omega_N t\right)\right)\ket{\downarrow_X}\bra{\downarrow_X}\right.\\\nonumber
&\left.+i\left(2p-1\right)\sin\left(\omega_N t\right)e^{2iG_j}\ket{\uparrow_X}\bra{\downarrow_X}-i\left(2p-1\right)\sin\left(\omega_N t\right)e^{-2iG_j}\ket{\downarrow_X}\bra{\uparrow_X}\right]^{\otimes N}
\end{align*}
\begin{align*}
+\frac{1}{4}\ket{\uparrow_Z}\bra{\downarrow_Z}_{S}&\prod_{j=1}^N\otimes\left[\left(1+\left(2p-1\right)\cos\left(\omega_N t\right)\right)e^{-2iG_j}\ket{\uparrow_X}\bra{\uparrow_X}+\left(1-\left(2p-1\right)\cos\left(\omega_N t\right)\right)e^{2iG_j}\ket{\downarrow_X}\bra{\downarrow_X}\right.\\\nonumber
&\left.+i\left(2p-1\right)\sin\left(\omega_N t\right)\ket{\uparrow_X}\bra{\downarrow_X}-i\left(2p-1\right)\sin\left(\omega_N t\right)\ket{\downarrow_X}\bra{\uparrow_X}\right]^{\otimes N}
\end{align*}
\begin{align*}
+\frac{1}{4}\ket{\downarrow_Z}\bra{\uparrow_Z}_{S}\prod_{j=1}^N\otimes&\left[\left(1+\left(2p-1\right)\cos\left(\omega_N t\right)\right)e^{2iG_j}\ket{\uparrow_X}\bra{\uparrow_X}+\left(1-\left(2p-1\right)\cos\left(\omega_N t\right)\right)e^{-2iG_j}\ket{\downarrow_X}\bra{\downarrow_X}\right.\\\nonumber
&\left.+i\left(2p-1\right)\sin\left(\omega_N t\right)\ket{\uparrow_X}\bra{\downarrow_X}-i\left(2p-1\right)\sin\left(\omega_N t\right)\ket{\downarrow_X}\bra{\uparrow_X}\right]^{\otimes N}.
\end{align*}
The reduced density matrix of the sensor, given by tracing out the nuclei in eq. \eqref{pol3},
\begin{align}\label{pol5}
\rho_{NV}=\frac{1}{2}\left[\ket{\uparrow_Z}\bra{\uparrow_Z}+\ket{\downarrow_Z}\bra{\downarrow_Z}+\ket{\uparrow_Z}\bra{\downarrow_Z}\prod_{j=1}^N\left(\cos\left(2G_j\right)-i\left(2p-1\right)\sin\left(2G_j\right)\cos\left(\omega_N t\right)\right)+\textrm{h.c.}\right].
\end{align}
As in previous sections, we rewrite the off-diagonal elements of the density matrix \eqref{pol5} in polar representation
\begin{align}\label{pol6}
&\Phi_j=\arctan\left[\left(2p-1\right)\cos\left(\omega_N t\right)\tan\left(2G_j\right)\right],\\\label{pol7}
&r_j=\sqrt{\cos^2\left(2 G_j\right)+\left(2p-1\right)^2\sin^2\left(2G_j\right)\cos^2\left(\omega_N t\right)}=\sqrt{1-\left[1-\left(2p-1\right)^2\cos^2\left(\omega_N t\right)\right]\sin^2\left(2G_j\right)}\\\nonumber
&=\sqrt{1-\left[\left(1-\left(2p-1\right)^2\right)+\left(2p-1\right)^2\sin^2\left(\omega_N t\right)\right]\sin^2\left(2G_j\right)}=\sqrt{1-\left[4p\left(1-p\right)+\left(2p-1\right)^2\sin^2\left(\omega_N t\right)\right]\sin^2\left(2G_j\right)}.
\end{align}
Eqs. \eqref{pol6} and \eqref{pol7} are identical to eqs. \eqref{spatial5} and \eqref{spatial6} respectively when $p=0$ or $p=1$, as expected.
Eq. \eqref{pol6}, hence shows, that the signal is equal to the one in a fully polarized ensemble multiplied by the polarization, $\textrm{pol}\equiv2p-1$, in the weak coupling limit $G_j\ll1$. The decay \eqref{pol7}, however, consists of a two terms - one that can be associated with the polarized dynamics and the other with the unpolarized dynamics. The first, is the same term as the in the fully polarized case, multiplied by the polarization squared, $\textrm{pol}^2=(2p-1)^2$. The other is a constant term proportional to the Bernoulli variance, which associated with the classical dephasing.
The dynamics can be thought in the following way - the polarized spins follow the protocol as before, while the unpolarized spins create only classical magnetic field fluctuations which makes the sensor dephase. The calculation of the mean magnetic field and $B_{rms}$ is the same as in sec. \ref{Spatially_dependent}.
The QFI for times $\tau\ll\tau_D$,
\beq\label{pol8}
\pazocal{I}=4 \gamma_e^2\tau^2t^2 \sin ^2(\theta )\textrm{pol}^2 e^{-4 \gamma _e^2 B_{rms}^2 \tau ^2  \left[\left(1-\textrm{pol}^2\right)+\textrm{pol}^2\sin^2\left(\theta\right)\right]}\left[\left<B\right>^2+B_{rms}^2\frac{4  \gamma _e^2 B_{rms}^2 \tau ^2   \cos ^2(\theta)}{1-e^{-4 \gamma _e^2 B_{\text{rms}}^2  \tau ^2\left[\left(1-\textrm{pol}^2\right)+\textrm{pol}^2\sin^2\left(\theta\right)\right]}}\right],
\eeq
where $\left<B\right>$ and $B_{rms}$ are the ones of a fully polarized ensemble given by eqs. \eqref{spatial9} and \eqref{spatial12}. In the weak back-action limit, $\gamma_eB_{rms}\tau\ll1$ , the optimal QFI is given for $\sin^2\theta=1$, for which,
\beq\label{pol9}
\pazocal{I}=4pol^2\left(\gamma_e\left<B\right>\tau\right)^2 t^2.
\eeq
Eq. \eqref{pol9} is identical to \eqref{spatial21} up to a reduction by a factor of $pol^2$, which is expected as $\left<B\right>_{\textrm{partialy\ polarized}}=pol\cdot\left<B\right>_{\textrm{fully\ polarized}}$. In the strong back-action limit, we can differ to the two strategies of sec. \ref{undriven}, choosing
\begin{align}\label{pol10}
\tau_1^2=\frac{1}{4 \gamma _e^2 B_{rms}^2 \tau ^2 },\sin^2\left(\theta\right)=1,
\end{align}
or
\begin{align}\label{pol11}
\tau_2^2=\frac{1}{4 \gamma_e^2 B_{rms}^2 \textrm{pol}^2\sin^2\left(\theta\right)}
\end{align}
Eq. \eqref{pol10} leads to the QFI,
\beq
\pazocal{I}_1=\frac{t^2}{e}\left(\frac{pol\cdot\left<B\right>}{B_{rms}}\right)^2,
\eeq
while eq. \eqref{pol11} leads to the QFI 
\beq
\pazocal{I}_2=\frac{t^2}{e}\left(\frac{\left<B\right>}{B_{rms}}\right)^2e^{-4 \gamma_e^2 B_{rms}^2 \tau ^2 (1-\textrm{pol}^2)}.
\eeq
\section{Multiple NVs - constant coupling}

Finally, we extend our protocol to consider multiple sensors. Given $M$ sensors the free Hamiltonian in the interaction picture with respect to their energy gap is
\beq\label{multi1}
H_0=\frac{\omega_N}{2}\sum_{j=1}^NI_z^j.
\eeq
 The simplified interaction Hamiltonian is
\beq\label{multi2}
H_1=g\sum_{i=1}^N\sum_{j=1}^M\sigma_z^jI_x^i
\eeq
The system is initialized to the state
\beq\label{multi3}
\ket{\uparrow_X}_S^{\otimes M}\otimes\ket{\uparrow_X}^{\otimes N}.
\eeq
The first step of the protocol, propagation of the nuclei for a time $t$ according to the free Hamiltonian, remains the same,
\beq\label{multi4}
\ket{\psi_t}=\ket{\uparrow_X}^{\otimes M}\left(\cos\left(\frac{\theta}{2}\right)\ket{\uparrow_X}-i\sin\left(\frac{\theta}{2}\right)\ket{\downarrow_X}\right)^{\otimes N}.
\eeq
The main difference is that the second rotation, given by the propagation of \eqref{multi4} according to the interaction \eqref{multi2}, is now affected by the total $Z$ spin of the sensors,
\beq\label{multi5}
\ket{\psi_\tau}=\frac{1}{\sqrt{2^M}}\sum_{\bar{x}\in\{0,1\}^M}\ket{\bar{x}}_S\left(\cos\left(\frac{\theta}{2}\right) e^{i 2 g \tau \sum_i x_i}\ket{\uparrow_X}-i e^{-2 i g \tau \sum_i x_i}\sin\left(\frac{\theta}{2}\right)\ket{\downarrow_X}\right)^{\otimes N},
\eeq
where we substituted $\ket{\uparrow_X}_S,\ket{\downarrow_X}_S$ by $\ket{0}_S,\ket{1}_S$ for brevity.
The density matrix of \eqref{multi5} is
\begin{align}\label{multi6}
&\rho_\tau
=\frac{1}{2^M}\sum_{\bar{x},\bar{y}\in\{0,1\}^M}\ket{\bar{x}}\bra{\bar{y}}\left(\cos^2\left(\frac{\theta}{2}\right) e^{i 2 g \tau \left(\sum_i x_i-\sum_i y_i\right)}\ket{\uparrow_X}\bra{\uparrow_X}+ e^{- i 2 g \tau \left(\sum_i x_i-\sum_i y_i\right)}\sin^2\left(\frac{\theta}{2}\right)\ket{\downarrow_X}\bra{\downarrow_X}+...\right)^{\otimes N}.
\end{align}
The reduced density operator of the sensors is thus
\begin{align}\label{multi7}
&\rho_\tau=\frac{1}{2^M}\sum_{\bar{x},\bar{y}\in\{0,1\}^M}\ket{\bar{x}}\bra{\bar{y}}\left[\cos^2\left(\frac{\theta}{2}\right) e^{i 2 g \tau \left(\sum_i x_i-\sum_i y_i\right)}+ e^{- i 2 g \tau \left(\sum_i x_i-\sum_i y_i\right)}\sin^2\left(\frac{\theta}{2}\right)\right]^N\\\nonumber
&=\frac{1}{2^M}\sum_{\bar{x},\bar{y}\in\{0,1\}^M}\ket{\bar{x}}\bra{\bar{y}}\left(\cos\left[2 g\tau\left(\sum_i x_i-\sum_i y_i\right)\right]+i\sin\left[2 g\tau\left(\sum_i x_i-\sum_i y_i\right)\right]\cos\theta\right)^N.
\end{align}

\begin{figure*}[t]
	\subfloat[]{\includegraphics[width=0.48\textwidth]{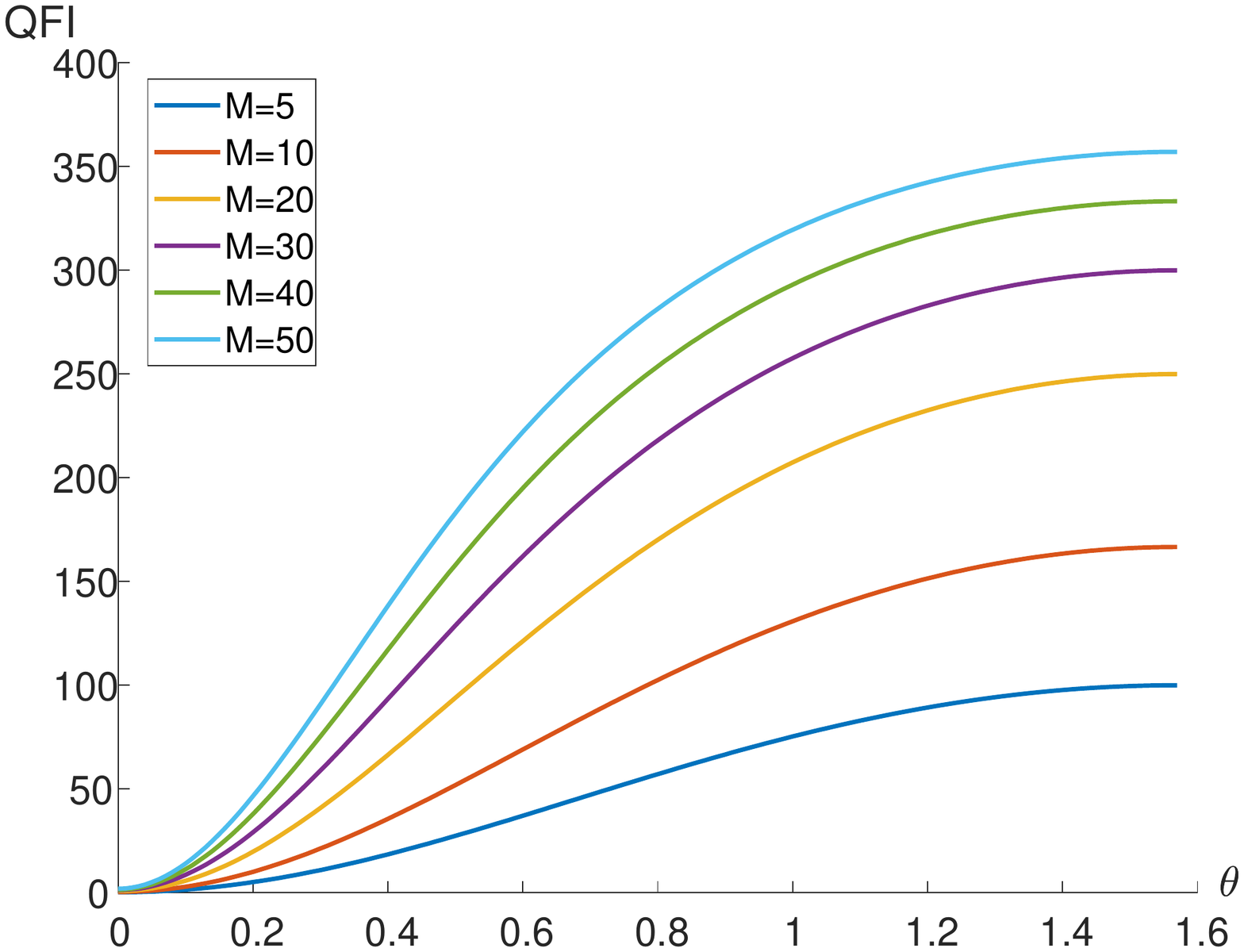} \label{QFI_multi_weak}}
	\hfill
	\subfloat[]{\includegraphics[width=0.48\textwidth]{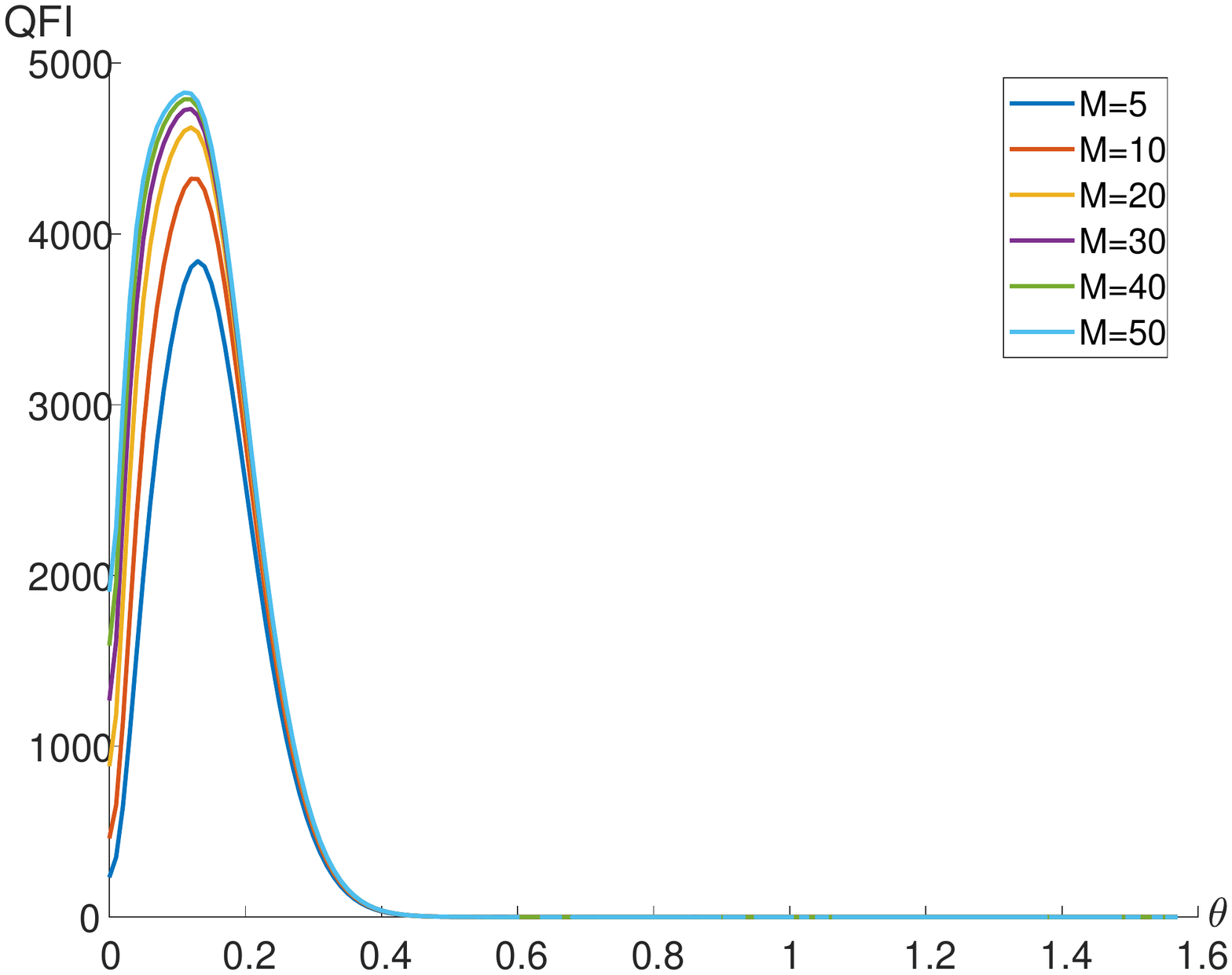} \label{QFI_multi_strong}}
	\hfill
	\caption{(a) The QFI estimated numerically by eq. \eqref{QFI_numerics} for different numbers of NVs with $N=500$ and $g\tau=0.01$ resulting in weak back-action. The optimal QFI is achieved for $\theta=\frac{pi}{2}$ as the single NV case. More NVs result in a higher QFI, which is still far from the SQL. (b) The QFI estimated numerically by eq. \eqref{QFI_numerics} for different numbers of NVs with $N=5000$ and $g\tau=0.1$ resulting in strong back-action. The optima $\theta$ remains the same, regardless of the number of NVs. As the number of NVs grows the QFI becomes wider approaches the SQL.}
	\label{QFI_multi} 
\end{figure*}

So in general, the density matrix reads
\begin{equation}\label{multi12}
\rho=\frac{1}{2^{M}}  \sum_{\bar{x},\bar{y}\in\{0,1\}^M} r\left(\bar{x}-\bar{y}\right)e^{i\Phi\left(\bar{x}-\bar{y}\right)}\ket{\bar{x}}\bra{\bar{y}}.
\end{equation}
Since there is a dependence only on the sum of $\bar{x},\bar{y}$, let us denote these sums as $s_{x},s_{y}$ and rewrite \eqref{multi12},

\begin{equation}\label{multi14}
\rho=\frac{1}{2^{M}} \sum_{s_x,s_y=0}^M r\left(s_{x}-s_{y}\right)e^{i\Phi\left(s_{1}-s_{2}\right)}\sqrt{n\left(s_{x}\right)}\sqrt{n\left(s_{y}\right)}\left(\frac{1}{\sqrt{n\left(s_{x}\right)}}\underset{x\in s_{y}}{\sum}|x\rangle\right)\left(\frac{1}{\sqrt{n\left(s_{y}\right)}}\underset{y\in s_{2}}{\sum}\langle y|\right).
\end{equation}
In the form \eqref{multi14}, it is clear that the matrix acts non-trivially only on symmetric states in the total angular momentum basis. Let $|s_{1}\rangle$ stand for the symmetric state of the sum $s_{1},$ then the density matrix reads
\begin{equation}\label{multi15}
\rho=\frac{1}{2^{M}} \sum_{s_x,s_y=0}^M r\left(s_{x}-s_{y}\right)e^{i\Phi\left(s_{1}-s_{2}\right)}\sqrt{n\left(s_{x}\right)}\sqrt{n\left(s_{y}\right)}\ket{s_x}\bra{s_y}.
\end{equation}
Hence, the density matrix is basically an $M+1$ by $M+1$ matrix (M  being the number of NVs) with:
\begin{equation}\label{multi16}
\rho_{j,k}=\frac{1}{2^{M}} \left(\cos\left(2g\tau\left(j-k\right)\right)+i\sin\left(2g\tau\left(j-k\right)\right)\cos\left(\phi\right)\right)^{N}\sqrt{\binom{n}{j}}\sqrt{\binom{n}{k}}.
\end{equation}
The QFI for multiple NV's can be estimated numerically by the relation \cite{QFI}
\beq\label{QFI_numerics}
\pazocal{I}=8\frac{1-\textrm{Tr}\{\sqrt{\sqrt{\rho(\theta)}\rho(\theta+d\theta)\sqrt{\rho(\theta)}}\}}{d\theta^2}t^2.
\eeq
The results are depicted in fig. \ref{QFI_multi} - the number of NVs does not change the optimal $\theta$ and as the number of sensors increases the QFI approaches the SQL in the strong back-action regime.

\begin{figure*}[t]
	\subfloat[]{\includegraphics[width=0.48\textwidth]{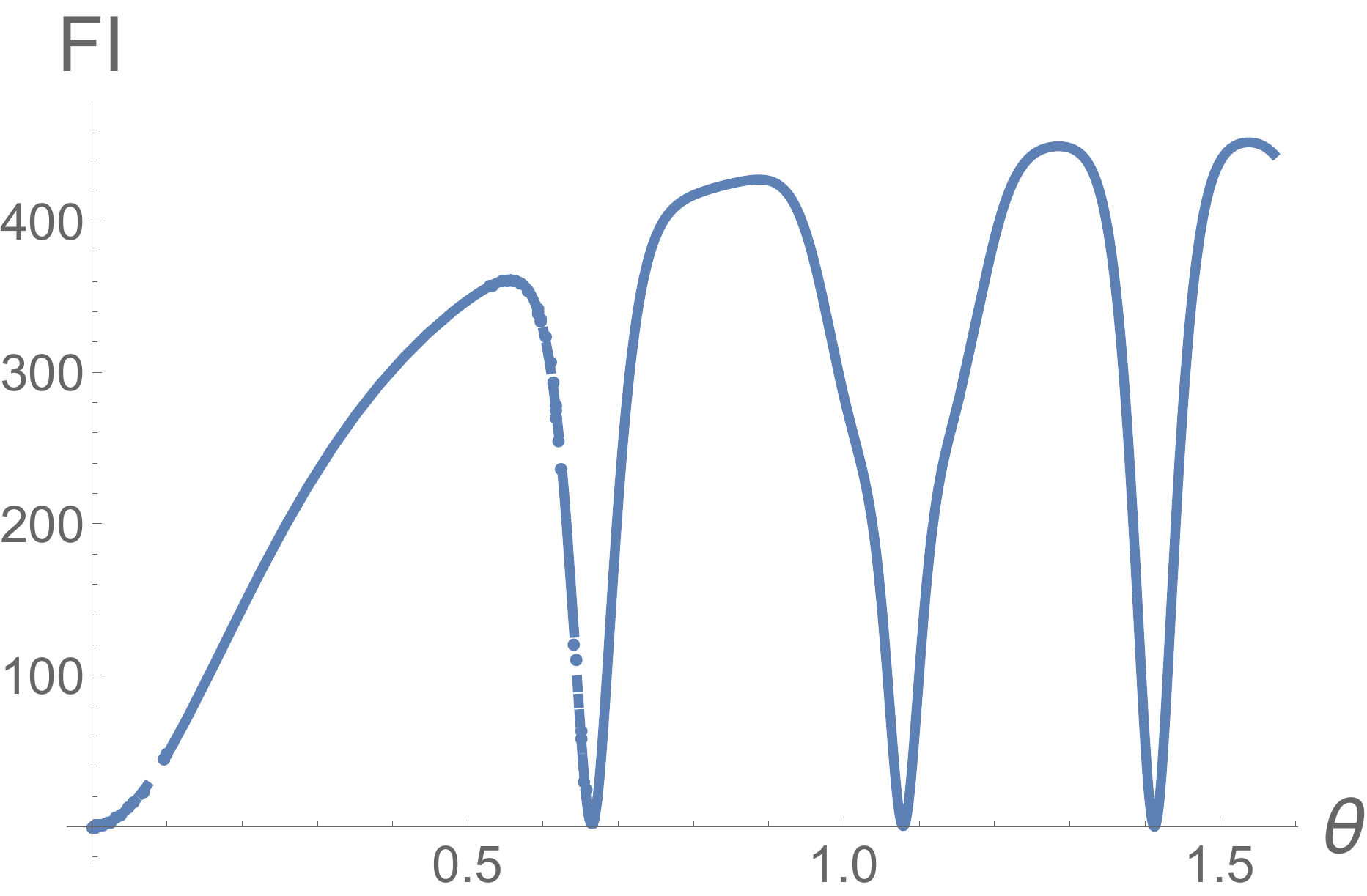} \label{FIY_weak}}
	\hfill
	\subfloat[]{\includegraphics[width=0.48\textwidth]{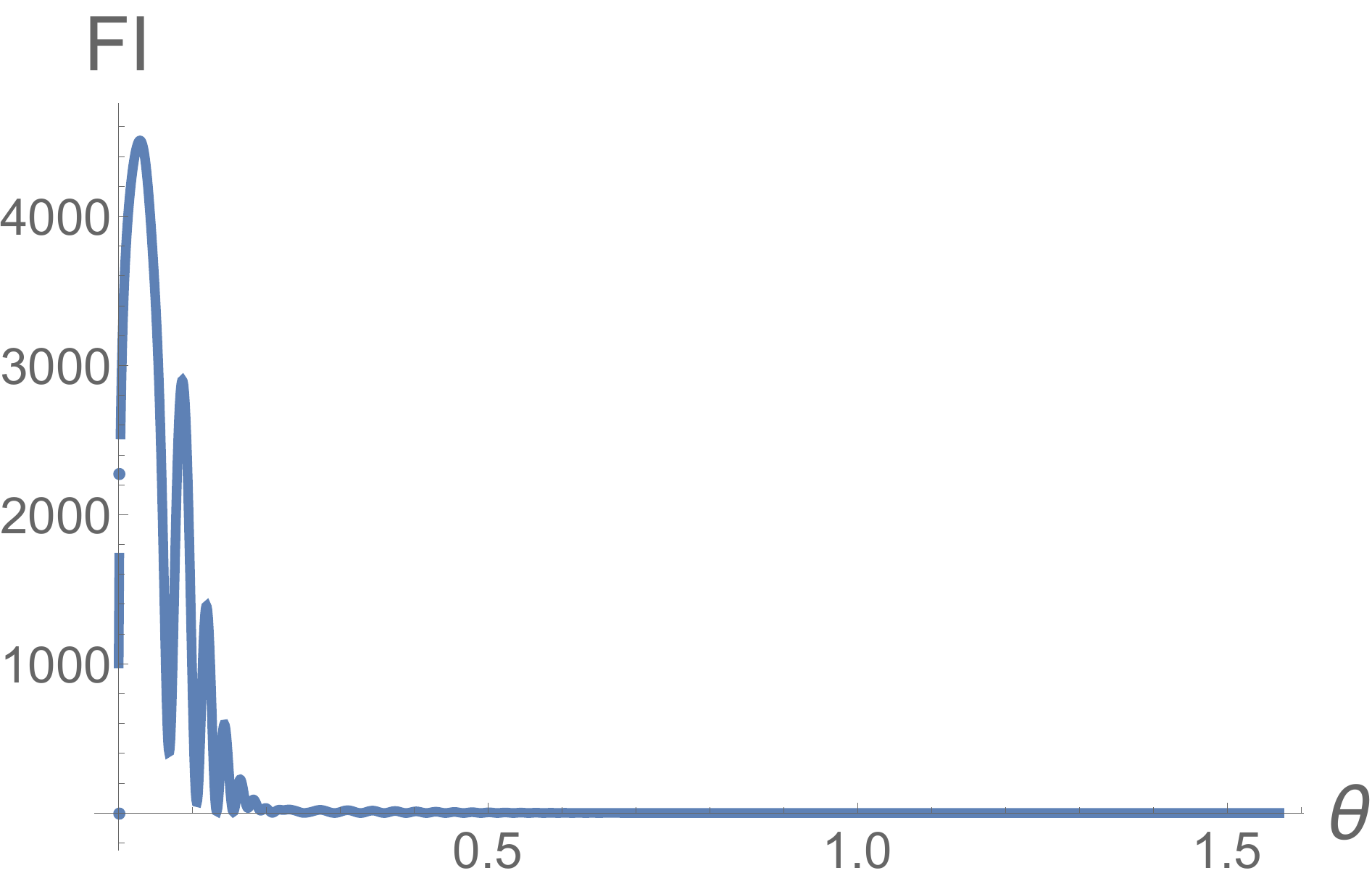} \label{FIY_strong}}
	\hfill
	\caption{(a) The FI estimated numerically by eq. \eqref{FI_Y} with $N=500$, $g\tau=0.01$ and $M=50$ resulting in weak back-action. The optimal QFI is achieved for $\theta$ slightly smaller than $\frac{\pi}{2}$ as predicted by \ref{QFI_multi_weak}. The value of the FI is in reasonable agreement with \ref{QFI_multi_weak}, which implies that the optimal measurement basis is not entangled. (b) The QFI estimated numerically by eq. \eqref{FI_Y} with $N=5000$, $g\tau=0.1$ and $M=50$ resulting in strong back-action. The optimal $\theta$ deviates strongly from \ref{QFI_multi_strong} and the FI does not reach it's optimum value. It is therefore not clear whether there exists an unentangled basis that can reproduce the QFI.}
	\label{FI_Y_numerics} 
\end{figure*}

The QFI demonstrates the promise of achieving the SQL using multiple sensors. However, the measurement basis can generally be highly entangled. To estimate whether the optimal basis is entangled, we would like to estimate the FI of a measurement in the $Y$ basis. It is easier to work in the form \eqref{multi7}, where the transformation to the $Y$ basis yields
\begin{align}
\rho_\tau=\frac{1}{2^{2M}}\sum_{\bar{x},\bar{y}\in\{0,1\}^M}&\sum_{\bar{a},\bar{b}\in\{0,1\}^M}\left(-i\right)^{\sum_i x_i-y_i}\left(-1\right)^{\bar{x}\cdot\bar{a}+\bar{y}\cdot\bar{b}}\ket{\bar{a}}\bra{\bar{b}}\times\\
\nonumber&\left(\cos\left[2g\tau\left(\sum_i x_i-\sum_i y_i\right)\right]+i\sin\left[2g\tau\left(\sum_i x_i-\sum_i y_i\right)\right]\cos\theta\right)^N.
\end{align}
Therefore, the probability to get a result $\bar{z}$ is
\beq\label{multi17}
P_{\bar{z}}=\frac{1}{2^{2M}}\sum_{\bar{x},\bar{y}\in\{0,1\}^M}\left(-i\right)^{s_x-s_y}\left(-1\right)^{\left( \bar{x}+\bar{y}\right)\cdot\bar{z}}\left(\cos\left[2g\tau\left(s_x-s_y\right)\right]+i\sin\left[2g\tau\left(s_x-s_y\right)\right]\cos\theta\right)^N.
\eeq
Eq. \eqref{multi17} can be written in terms of the sums
\begin{align}\label{multi18}
&P_{\bar{z}}=\frac{1}{2^{2M}}\sum_{\bar{x},\bar{y}\in\{0,1\}^M}\left(-i\right)^{s_x-s_y}a_{s_x,s_z} a_{s_y,s_z}\left(\cos\left[2g\tau\left(s_x-s_y\right)\right]+i\sin\left[2g\tau\left(s_x-s_y\right)\right]\cos\theta\right)^N\\\nonumber
&=\frac{1}{2^{2M}}\sum_{s_x,s_y=0}^M\left(-i\right)^{s_x-s_y}a_{s_x,s_z} a_{s_y,s_z}\left(\cos\left[2g\tau\left(s_x-s_y\right)\right]+i\sin\left[2g\tau\left(s_x-s_y\right)\right]\cos\phi\right)^N\\\nonumber
&=\frac{1}{2^{2M}}\sum_{s=0}^M  a_{s,s_z}^2+\frac{2}{2^{2M}}\sum_{s_x=0}^M\sum_{s_y=s_x+1}^M a_{s_x,s_z} a_{s_y,s_z}\textrm{Re}\left[\left(-i\right)^{s_x-s_y}\left(\cos\left[2g\tau\left(s_x-s_y\right)\right]+i\sin\left[2g\tau\left(s_x-s_y\right)\right]\cos\theta\right)^N\right]
\end{align}
where $s_h=\sum_i h_i$ and $a_{s,s_z}=\sum_{k=\max\{s+s_z-M\}}^{\min\{s,s_z\}}\left(-1\right)^k\binom{s_z}{k}\binom{M-s_z}{s-k}$.
The FI will be given by
\begin{align}\label{FI_Y}
\pazocal{F}=\sum_{\bar{z}\in\{0,1\}^M}\frac{\left(\frac{\partial P_{\bar{z}}}{\partial \omega_N}\right)^2}{P_{\bar{z}}}=\sum_{s_z=0}^M\frac{\left(\frac{\partial P_{s_z}}{\partial \omega_N}\right)^2}{P_{s_z}}.
\end{align}
The numerical evaluation of eq. \eqref{FI_Y} is presented in fig. \ref{FI_Y_numerics} - in the weak back-action regime it appears that the $Y$ basis is within reasonable agreement with the QFI, meaning it is close to the optimal measurement basis. When the back-action is strong, however, the optimal $\theta$ of the FI in the $Y$ basis ($\theta\approx0.03$) differs strongly from the one of the QFI ($\theta\approx0.19$), which results in a decreased optimum. Though it is not substantial, it is still unclear whether an unentangled measurement basis can achieve the QFI. 

%\textbf{Claim:} When $\phi=0$ the state $\ket{\psi}=\frac{1}{\sqrt{\sum_{s_n}{n \choose |s_n|}}} \sum_{s_n}\sqrt{n \choose |s_n|}e^{-ig\tau s_n}\ket{s_n}$ is an eigenstate of $\rho_{\tau}$.
%\textbf{Proof:} When $\phi=0$ the density matrix \eqref{multi15} has the simple form
%\beq
%\rho_\tau=\sum_{s_1,s_2}\sqrt{n \choose |s_1|}\sqrt{n \choose |s_2|}e^{-ig\tau\left(s_1-s_2\right)}\ket{s_1}\bra{s_2}.
%\eeq
%Applying $\rho$ to the state $\ket{\psi}$ will yield 
%\begin{align}
%&\rho_\tau\ket{\psi}=\sum_{s_1,s_n}\frac{1}{\sqrt{\sum_{s_n}{n \choose |s_n|}}}\sqrt{n \choose |s_n|}e^{-ig\tau s_n}\sqrt{n \choose |s_1|}\sqrt{n \choose |s_n|}e^{-ig\tau\left(s_1-s_n\right)}\ket{s_1}\\
%&=\frac{1}{\sqrt{\sum_{s_n}{n \choose |s_n|}}}\sum_{s_n}{n \choose |s_n|}\sum_{s_1}\sqrt{n \choose |s_1|}e^{-ig\tau s_1}\ket{s_1}=\ket{\psi}
%\end{align}
%%%%%%%%%%%%%%%%%%%%%%%%%%%%%%%%%
     
\appendix
\section{Angular momentum decomposition of the dipole-dipole interaction}
\label{Appendix: DD}
In following we show that the diople-dipole interaction Hamiltonian of two spins $\bar{S}$ and $\bar{I}$,
\beq\label{DDHamiltonian}
H_{DD}/J=-\left|\bar{r}\right|^{-3}\left(3\left(\bar{S}\cdot\hat{r}\right)\left(\bar{I}\cdot\hat{r}\right)-\bar{S}\cdot\bar{I}\right),
\eeq
with  $J=\frac{\mu_0 \gamma_{NV}\gamma_{B}}{4\pi}=0.49 \textrm{MHz}\cdot\textrm{nm}^3$,
 can be decomposed into six parts associated with conservation of angular momentum \cite{Abragam1961},
\begin{align}\label{Decomposition}
&A=-4\sqrt{\frac{\pi}{5}}\left|\bar{r}\right|^{-3}Y_2^0(\Omega)S_zI_z,\ B=-\frac{1}{4} \left(4\sqrt{\frac{\pi}{5}}\right)\left|\bar{r}\right|^{-3}\left(S_+I_-+S_-I_+\right)Y_2^0(\Omega), \\\nonumber
&C=-\frac{3}{2}\left|\bar{r}\right|^{-3}\left(2\sqrt{(\frac{2\pi}{15})}\right)\left(S_zI_++I_zS_+\right)Y_2^{-1}(\Omega),\
D=\frac{3}{2}\left|\bar{r}\right|^{-3}\left(2\sqrt{(\frac{2\pi}{15})}\right)\left(S_zI_-+I_zS_-\right)Y_2^{1}(\Omega),\\\nonumber
&E=-\frac{3}{4}\left(4\sqrt{(\frac{2\pi}{15})}\right)Y_2^{-2}(\Omega)S_+I_+,\
F=-\frac{3}{4}\left(4\sqrt{(\frac{2\pi}{15})}\right)Y_2^{2}(\Omega)S_-I_-,
\end{align}
where $H_{DD}/J=A+B+C+D+E+F$ and $Y_l^m(\Omega)$ are the spherical harmonics.
We start by writing the inner product in \eqref{DDHamiltonian} explicitly,
\begin{align}\label{DD1}
H_{DD}/J=-&\left|\bar{r}\right|^{-3}\left(3\left(S_x\sin(\theta)\cos(\varphi)+S_y\sin(\theta)\sin(\varphi)+S_z\cos(\theta)\right)\times\right.\\
\nonumber&\left.\left(I_x\sin(\theta)\cos(\varphi)+I_y\sin(\theta)\sin(\varphi)+I_z\cos(\theta)\right)-S_xI_x-S_yI_y-S_zI_z\right)
\end{align}
The product of the two z components will yield
\beq
A=-\left|\bar{r}\right|^{-3}\left(3\cos^2(\theta)-1\right)S_zI_z
\eeq
which is equivalent to \eqref{Decomposition} using the definition $Y_2^0\left(\Omega\right)=\frac{1}{4}\sqrt{\frac{5}{\pi}}\left(3\cos^2\theta-1\right)$.
The sum of the terms proportional to $S_xI_x$ and $S_yI_y$ in \eqref{DD1} can be rewritten with the spin lowering and raising operators $S_\pm=S_x\pm iS_y,\ I_\pm=I_x\pm i I_y$
\begin{align}\label{DD2}
-&\left|\bar{r}\right|^{-3}\left[\left(3\sin^2(\theta)\cos^2(\varphi)-1\right)S_xI_x+\left(3\sin^2(\theta)\sin^2(\varphi)-1\right)S_yI_y\right]\\
&=-\frac{1}{4}\left|\bar{r}\right|^{-3}\left[\left(3\sin^2(\theta)\cos^2(\varphi)-1\right)(S_++S_-)(I_++I_-)-\left(3\sin^2(\theta)\sin^2(\varphi)-1\right)(S_+-S_-)(I_+-I_-)\right]\\\label{DD10}
&=-\frac{1}{4}\left|\bar{r}\right|^{-3}\left[\left(S_+I_-+S_-I_+\right)\left(3\cos^2(\theta)-1\right)+3\left(S_+I_++S_-I_-\right)\sin^2(\theta)\cos(2\varphi)\right].
\end{align}
The first contribution in \eqref{DD2},
\beq
B=-\frac{1}{4} \left(4\sqrt{\frac{\pi}{5}}\right)\left|\bar{r}\right|^{-3}\left(S_+I_-+S_-I_+\right)Y_2^0(\Omega)
\eeq
is the second term in \eqref{Decomposition}.
The terms $A$ and $B$ do not change the total angular momentum of the two spins, therefore, they are proportional to the spherical harmonic with $m=0$.
The terms proportional to$S_xI_y$ $S_YI_x$ in \eqref{DD1} yield similarly 
\begin{align}
-&3\left|\bar{r}\right|^{-3}\left(S_xI_y+S_yI_x\right)\sin^2(\theta)\sin(\varphi)\cos(\varphi)\\\label{DD3}
&=\frac{3}{4}\left|\bar{r}\right|^{-3} i \left[S_+I_+-S_-I_-\right]\sin^2(\theta)\sin(2\varphi).
\end{align}
Summing \eqref{DD3} with the remaining contribution of \eqref{DD10} results in the terms $E$ and $F$ in \eqref{Decomposition} with the definition $Y_2^{\pm2}=\frac{1}{4}\sqrt{\frac{15}{2\pi}}e^{\pm2i\varphi}\sin^2\theta$. The two operators $E/F$ raise/lower the angular momentum by two quanta, and therefore together with the spherical harmonic with $m=-2/2$ they conserve angular momentum. The remaining elements of \eqref{DD1} can be rewritten as
\begin{align}
-&3\left|\bar{r}\right|^{-3}\left[S_z\left(I_x\cos(\varphi)+I_y\sin(\varphi)\right)+I_z\left(S_x\cos(\varphi)+S_y\sin(\varphi)\right)\right]\cos(\theta)\sin(\theta)\\
-&\frac{3}{2}\left|\bar{r}\right|^{-3}\left[\left(S_zI_++I_zS_+\right)e^{-i\varphi}+\left(S_zI_-+I_zS_-\right)e^{i\varphi}\right]\cos(\theta)\sin(\theta),
\end{align}
which using the definition $Y_2^{\pm1}\left(\Omega\right)=\mp \frac{1}{2}\sqrt{\frac{15}{2\pi}}e^{\pm i \varphi}\cos\theta \sin\theta$ leads to the terms $C$ and $D$ in \eqref{Decomposition}. 

For reasons of convenience, in the next sections we use the notations 
\begin{align}
\label{SH_coeff}
&\zeta_0=-1,\ \tilde{\zeta}_0=-4\sqrt{\frac{\pi}{5}}\\\nonumber
&\zeta_1=\frac{3}{2},\ \tilde{\zeta}_1=\frac{3}{2}\left(2\sqrt{\frac{2\pi}{15}}\right)\\\nonumber
&\zeta_2=-\frac{3}{4},\ \tilde{\zeta}_2=-\frac{3}{4}\left(4\sqrt{\frac{2\pi}{15}}\right)\\\nonumber
&\zeta_{-m}= \zeta_m,\ 	\tilde{\zeta}_{-m}=\left(-1\right)^m \tilde{\zeta}_m.
\end{align}

\section{Calculating the mean field}\label{Mean_field}
\label{Sec_mean_planar}
In the following section we shall calculate the mean field integral in a planar geometry, meaning, in a setting where the NV is situated in a depth $d$  below a planar diamond surface and the nuclear spins occupy the entire half-space above the diamond. The result of the section is
\beq\label{mean_field_planar}
\gamma_e\left<B\right>=\pi J \sin\left(2\alpha\right),
\eeq  
The calculation is carried out in a spherical coordinate system, where the NV is found at the origin and the $\hat{z}$ axis is perpendicular to the diamond surface.

The mean field at the NV's location is a function of the following integrals,
\beq\label{mean_field_int}
I_{1}^{\left(m\right)}=\tilde{\zeta}_m\int d^3r \frac{Y_2^{\left(m\right)}}{r^3},
\eeq
where $\tilde{\zeta}_m$ are defined by \eqref{SH_coeff}.
In the given geometry $\varphi\in\left[0,2\pi\right],\ \theta\in\left[0,\frac{\pi}{2}\right],\ r\in\left[\frac{d}{\cos\theta},\infty\right]$, therefore,
\beq\label{mean_field_int2}
I_{1}^{{\left(m\right)}}=\tilde{\zeta}_m\int\limits_{0}^{2\pi} d\varphi\int\limits_{0}^{1} d\cos\theta Y_2^{\left(m\right)}\left(\Omega\right) \int\limits_{d/\cos\theta}^{r_{co}} \frac{dr}{r}=\tilde{\zeta}_m\int\limits_{0}^{2\pi} d\varphi\int\limits_{0}^{1} d\cos\theta Y_2^{\left(m\right)}\left(\Omega\right) \ln\left(\frac{r_{co}}{d}\cos\theta\right),
\eeq
where $r_{co}$ is some cut off radius, that shouldn't effect the final result.
For a non-tilted NV, meaning, when the NV's magnetization axis coincides with the $\hat{z}$ axis, the integration over $\phi$ will eliminate the total contribution for $m\neq0$ due to the cylindrical symmetry. The remaining integral
\beq
\label{non_tilted_infinite}
I_{1}^{\left(0\right)}=2\pi\zeta_0\int\limits_{0}^{1} d\cos\theta (3\cos^2\theta-1) \ln\left(\frac{r_{co}}{d}\cos\theta\right)=-\frac{4\pi}{3}.
\eeq
If the NV is tilted the mean field can be calculated using the spherical harmonics rotation property,
\beq\label{rotation_identity}
Y_{2}^{\left(m\right)}\left(\bar{r}\right)=\sum_{m'=-2}^{2}\left[D_{mm'}^{\left(2\right)}\left(\mathcal{R}\right)\right]^{*}Y_{2}^{\left(m'\right)}\left(\bar{r}'\right),\ \bar{r}'=\mathcal{R}\bar{r},
\eeq 
where $D_{mm'}^{\left(2\right)}\left(\mathcal{R}\right)$ are the Wigner matrices with rotation $\mathcal{R}$. In our case $\bar{r}$ in \eqref{rotation_identity} are the coordinates in a system whose $\hat{z}$ axis is aligned with the NV's magnetization axis and $\bar{r}'$ are the coordinates in a rotated system whose $\hat{z}$ axis is perpendicular to the diamond surface.
Following the same argument as before, the integral \eqref{mean_field_int2} is non-zero only for $m'=0$,
\begin{align}
I_{1}^{\left(m\right)}&=2\pi\tilde{\zeta}_m\int\limits_{0}^{1} d\cos\theta \sum_{m'=-2}^{2}\left[D_{mm'}^{\left(2\right)}\left(\mathcal{R}\right)\right]^{*}Y_{2}^{\left(m'\right)}\left(\Omega\right) \ln\left(\frac{r_{co}}{d}\cos\theta\right)\\
&=2\pi\tilde{\zeta}_m\left[D_{m0}^{\left(2\right)}\left(\mathcal{R}\right)\right]^{*} \int\limits_{0}^{1} d\cos\theta Y_{2}^{\left(0\right)}\left(\Omega\right) \ln\left(\frac{r_{co}}{d}\cos\theta\right)
=-\frac{4\pi}{3}\frac{\tilde{\zeta}_m}{\tilde{\zeta}_0}\left[D_{m0}^{\left(2\right)}\left(\mathcal{R}\right)\right]^{*},
\end{align}
where in the last equality we used eq. \eqref{non_tilted_infinite}.
The rotation $\mathcal{R}$ of the NV is by an angle $\alpha$ around the Y axis, hence, the Wigner matrix has a simple form 
\begin{align}
I_{1}^{\left(1\right)}=\frac{\pi}{2}\sin\left(2\alpha\right)=I_{1}^{\left(-1\right)}
\end{align} 

In a scenario where only the $S_zI_x$ part of the dipole-dipole interaction is present, the mean field will be
\beq
 \left<B\right>=\frac{\hbar\mu_0\gamma_N}{4\pi}n\left(I_{1}^{\left(1\right)}+I_{-1}^{\left(1\right)}\right)=\frac{\hbar\mu_0\gamma_N}{4}n\sin\left(2\alpha\right).
\eeq
For a tilting angle of $\alpha=54.7$ and the number density of water $n=33\ \textrm{nm}^{-3}$,
\beq
\gamma_e\left<B\right>\approx48 \textrm{MHz}
\eeq

\section{Calculating $B_{rms}^2$ }\label{Decay_Planar}
In following section we would like to calculate $B_{rms}^2$ in a planar geometry as described in \ref{Sec_mean_planar}. Similar to the mean field, it is a function of the integrals 
\beq\label{decay_planar_int}
I_{2}^{\left(m_1,m_2\right)}=\tilde{\zeta}_{m_1}\tilde{\zeta}_{m_2}\int d^3r \frac{Y_2^{\left(m_1\right)}}{r^3}\frac{Y_2^{\left(m_2\right)}}{r^3}.
\eeq
These arise from the instantaneous limit ($t\rightarrow0)$ of the more general form,
\beq\label{decay_int}
I_{2}^{\left(m_1,m_2\right)}\left(t\right)=\tilde{\zeta}_{m_1}\tilde{\zeta}_{m_2}\int d^3r\int d^3r_0 \frac{Y_2^{\left(m_1\right)}\left(\Omega\right)}{r^3}\frac{Y_2^{\left(m_2\right)}\left(\Omega_0\right)}{r_0^3}P\left(\bar{r},\bar{r}_0,t\right),
\eeq
where $P\left(\bar{r},\bar{r}_0,t\right)$ is the diffusion propagator in the appropriate geometry.
In the following we calculate \eqref{decay_planar_int} in a planar geometry and show that
\beq\label{BRMS_planar}
\left<B^2_{rms}\right>=2\left(\frac{\hbar\mu_0\gamma_N}{4\pi}\right)^2 n\left(I_{2}^{\left(1,1\right)}+I_{2}^{\left(-1,1\right)}\right)=n\left(\frac{\hbar\mu_0\gamma_N}{4\pi}\right)^2 \frac{35 \pi -3 \pi  \cos (4 \alpha )}{256 d^3}.
\eeq
We start with the radial integration of \eqref{decay_planar_int},
\begin{align}\label{decay_planar_int2}
I_{2}^{\left(m_1,m_2\right)}&=\tilde{\zeta}_{m_1}\tilde{\zeta}_{m_2}\int d^3r \frac{Y_2^{\left(m_1\right)}}{r^3}\frac{Y_2^{\left(m_2\right)}}{r^3}=\tilde{\zeta}_{m_1}\tilde{\zeta}_{m_2}\int\limits_{0}^{2\pi} d\varphi\int\limits_{0}^{1} d\cos\theta Y_2^{\left(m_1\right)}\left(\Omega\right)Y_2^{\left(m_2\right)}\left(\Omega\right) \int\limits_{d/\cos\theta}^{\infty} \frac{dr}{r^4}\\
&=\frac{\tilde{\zeta}_{m_1}\tilde{\zeta}_{m_2}}{3}\int\limits_{0}^{2\pi} d\varphi\int\limits_{0}^{1} d\cos\theta Y_2^{\left(m_1\right)}\left(\Omega\right)Y_2^{\left(m_2\right)}\left(\Omega\right)\left(\frac{\cos\theta}{d}\right)^3.
\end{align}
Assuming the NV is not tilted, there will be non-zero contributions only when $m_2=-m_1$ due to the cylindrical symmetry,
\begin{align}
I_{2}^{\left(0,0\right)}&=\frac{2\pi}{3 d^3}\int\limits_{0}^{1} d\cos\theta \left(3\cos^2\theta-1\right)^2\cos^3\theta=\frac{\pi}{4 d^3}.\\
I_{2}^{\left(\pm 1,\mp 1\right)}&=\frac{3\pi}{2 d^3}\int\limits_{0}^{1} d\cos\theta \cos^5\theta\sin^2\theta=\frac{\pi}{16 d^3}.\\
I_{2}^{\left(\pm 2,\mp 2\right)}&=\frac{3\pi}{8 d^3}\int\limits_{0}^{1} d\cos\theta \cos^3\theta\sin^4\theta=\frac{\pi}{64 d^3}.\\
\end{align}

As in the previous sections, when the NV is tilted the spherical harmonics can be rotated using \eqref{rotation_identity},
\beq
I_{2}^{\left(m_1,m_2\right)}=\sum_{m_1',m_2'}\left[D_{m_1m_1'}^{\left(2\right)}\left(\mathcal{R}\right)\right]^{*}\left[D_{m_2m_2'}^{\left(2\right)}\left(\mathcal{R}\right)\right]^{*}\frac{\tilde{\zeta}_{m_1}\tilde{\zeta}_{m_2}}{3}\int\limits_{0}^{2\pi} d\varphi\int\limits_{0}^{1} d\cos\theta Y_2^{\left(m_1'\right)}\left(\Omega\right)Y_2^{\left(m_2'\right)}\left(\Omega\right)\left(\frac{\cos\theta}{d}\right)^3.
\eeq
The polar symmetry will enforce $m_1'=-m_2'$,
\beq
\label{decay1}
I_{2}^{\left(m_1,m_2\right)}=\sum_{m'}\left[D_{m_1 m'}^{\left(2\right)}\left(\mathcal{R}\right)\right]^{*}\left[D_{m_2-m'}^{\left(2\right)}\left(\mathcal{R}\right)\right]^{*}\frac{\tilde{\zeta}_{m_1}\tilde{\zeta}_{m_2}}{3}\int\limits_{0}^{2\pi} d\varphi\int\limits_{0}^{1} d\cos\theta Y_2^{\left(m\right)}\left(\Omega\right)Y_2^{\left(-m\right)}\left(\Omega\right)\left(\frac{\cos\theta}{d}\right)^3.
\eeq
Denoting the rotation angle as $\alpha$ we can rewrite \eqref{decay1} as
\beq
I_2^{\left(m_1,m_2\right)}\left(\alpha\right)=\sum_{m'}\left[D_{m_1 m'}^{\left(2\right)}\left(\mathcal{R}\right)\right]^{*}\left[D_{m_2-m'}^{\left(2\right)}\left(\mathcal{R}\right)\right]^{*}\frac{\tilde{\zeta}_{m_1}\tilde{\zeta}_{m_2}}{\tilde{\zeta}_{m'}\tilde{\zeta}_{-m'}}I_2^{\left(m',-m'\right)}\left(\alpha=0\right).
\eeq
In the special case, where the dipolar interaction is proportional to $S_zI_x$, only $m_1,m_2=\pm1$ are relevant, for which 
\begin{align}\label{decay_int_planar2}
I_2^{\left(1,1\right)}\left(\alpha\right)&=I_2^{\left(-1,-1\right)}\left(\alpha\right)=-\frac{3 (4 \pi  \cos (2 \alpha )+\pi  \cos (4 \alpha )-5 \pi )}{1024 d^3}\\\label{decay_int_planar3}
I_2^{\left(-1,1\right)}\left(\alpha\right)&=I_2^{\left(1,-1\right)}\left(\alpha\right)=\frac{12 \pi  \cos (2 \alpha )-3 \pi  \cos (4 \alpha )+55 \pi }{1024 d^3}
\end{align}
Eqs. \eqref{decay_int_planar2} and \eqref{decay_int_planar3} lead to \eqref{BRMS_planar}.

\section{Third order integrals}\label{Third_moment}
As the distance from the surface decreases, $B_{rms}\sim\left<B\right>$, and higher orders of the correlation function become important. 
These are generally hard to calculate, so here we shall focus on the instantaneous limit, where
\beq
\left(\int\limits_0^t \textrm{dt}'f'\left(t'\right)\right)^3\approx t^3 \sum_{m_1,m_2,m_3}C_{m_1,m_2,m_3}I_3^{\left(m_1,m_2,m_3\right)},
\eeq
where 
\beq\label{third_order1}
I_{3}^{\left(m_1,m_2,m_3\right)}=\tilde{\zeta}_{m_1}\tilde{\zeta}_{m_2}\tilde{\zeta}_{m_3}\int d^3r \frac{Y_2^{\left(m_1\right)}}{r^3}\frac{Y_2^{\left(m_2\right)}}{r^3}\frac{Y_2^{\left(m_3\right)}}{r^3}
\eeq
and $C_{m_1,m_2,m_3}$ are coefficients determined by NV's tilting and the interaction Hamiltonian.
In the following we calculate \eqref{third_order1},
\begin{align}
I_{3}^{\left(m_1,m_2,m_3\right)}&=\tilde{\zeta}_{m_1}\tilde{\zeta}_{m_2}\tilde{\zeta}_{m_3}\int d^3r \frac{Y_2^{\left(m_1\right)}}{r^3}\frac{Y_2^{\left(m_2\right)}}{r^3}\frac{Y_2^{\left(m_3\right)}}{r^3}=\tilde{\zeta}_{m_1}\tilde{\zeta}_{m_2}\tilde{\zeta}_{m_3}\int\limits_{0}^{2\pi} d\phi\int\limits_{0}^{1} d\cos\theta Y_2^{\left(m_1\right)}\left(\Omega\right)Y_2^{\left(m_2\right)}\left(\Omega\right)Y_2^{\left(m_3\right)}\left(\Omega\right) \int\limits_{d/\cos\theta}^{\infty} \frac{dr}{r^7}\\
&=\frac{\tilde{\zeta}_{m_1}\tilde{\zeta}_{m_2}\tilde{\zeta}_{m_3}}{6}\int\limits_{0}^{2\pi} d\phi\int\limits_{0}^{1} d\cos\theta Y_2^{\left(m_1\right)}\left(\Omega\right)Y_2^{\left(m_2\right)}\left(\Omega\right)Y_2^{\left(m_3\right)}\left(\Omega\right)\left(\frac{\cos\theta}{d}\right)^6.
\end{align}
For a non-tilted NV the only contribution will be from $m_1+m_2+m_3=0$, denoting $I_3^{\left(m_1.m_2,m_3\right)}\left(\alpha=0\right)\equiv J_3^{\left(m_1.m_2,m_3\right)}$,
\begin{align}
J_3^{\left(2,-2,0\right)}&=\frac{\tilde{\zeta}_{2}\tilde{\zeta}_{-2}\tilde{\zeta}_{0}}{6}\int\limits_{0}^{2\pi} d\varphi\int\limits_{0}^{1} d\cos\theta Y_2^{\left(2\right)}\left(\Omega\right)Y_2^{\left(-2\right)}\left(\Omega\right)Y_2^{\left(0\right)}\left(\Omega\right)\left(\frac{\cos\theta}{d}\right)^6
\\\nonumber
&=-\frac{3\pi}{16d^6}\int\limits_{0}^{1} d\cos\theta \sin^4\theta \left(3\cos^2\theta-1\right)\cos^6\theta=-\frac{4 \pi }{3003 d^6},
\end{align}
\begin{align}
J_3^{\left(1,1,-2\right)}&=\frac{\tilde{\zeta}_{1}\tilde{\zeta}_{1}\tilde{\zeta}_{-2}}{6}\int\limits_{0}^{2\pi} d\varphi\int\limits_{0}^{1} d\cos\theta Y_2^{\left(1\right)}\left(\Omega\right)Y_2^{\left(1\right)}\left(\Omega\right)Y_2^{\left(-2\right)}\left(\Omega\right)\left(\frac{\cos\theta}{d}\right)^6
\\\nonumber
&=-\frac{9\pi}{16d^6}\int\limits_{0}^{1} d\cos\theta \sin^4\theta\cos^8\theta=-\frac{\pi }{286 d^6},
\end{align}
\begin{align}
J_3^{\left(1,-1,0\right)}&=\frac{\tilde{\zeta}_{1}\tilde{\zeta}_{-1}\tilde{\zeta}_{0}}{6}\int\limits_{0}^{2\pi} d\varphi\int\limits_{0}^{1} d\cos\theta Y_2^{\left(1\right)}\left(\Omega\right)Y_2^{\left(-1\right)}\left(\Omega\right)Y_2^{\left(0\right)}\left(\Omega\right)\left(\frac{\cos\theta}{d}\right)^6
\\\nonumber
&=-\frac{3\pi}{4d^6}\int\limits_{0}^{1} d\cos\theta \sin^2\theta\cos^8\theta(3\cos^2\theta-1)=-\frac{7 \pi }{429 d^6},
\end{align}
\begin{align}
J_3^{\left(0,0,0\right)}&=\frac{\tilde{\zeta}_{0}\tilde{\zeta}_{0}\tilde{\zeta}_{0}}{6}\int\limits_{0}^{2\pi} d\varphi\int\limits_{0}^{1} d\cos\theta Y_2^{\left(0\right)}\left(\Omega\right)Y_2^{\left(0\right)}\left(\Omega\right)Y_2^{\left(0\right)}\left(\Omega\right)\left(\frac{\cos\theta}{d}\right)^6
\\\nonumber
&=-\frac{\pi}{3 d^6}\int\limits_{0}^{1} d\cos\theta \cos^6\theta\left(3\cos^2\theta-1\right)^3=-\frac{160 \pi }{1001 d^6}.
\end{align}
The tilted case is more complex since,
\begin{align}
I_{3}^{\left(m_1,m_2,m_3\right)}=\sum_{m_1',m_2',m_3'}&\left[D_{m_1m_1'}^{\left(2\right)}\left(\mathcal{R}\right)\right]^{*}\left[D_{m_2m_2'}^{\left(2\right)}\left(\mathcal{R}\right)\right]^{*}\left[D_{m_3m_3'}^{\left(2\right)}\left(\mathcal{R}\right)\right]^{*}\times\\\nonumber
&\frac{\tilde{\zeta}_{m_1}\tilde{\zeta}_{m_2}\tilde{\zeta}_{m_3}}{6}\int\limits_{0}^{2\pi} d\varphi\int\limits_{0}^{1} d\cos\theta Y_2^{\left(m_1'\right)}\left(\Omega\right)Y_2^{\left(m_2'\right)}\left(\Omega\right)Y_2^{\left(m_3'\right)}\left(\Omega\right)\left(\frac{\cos\theta}{d}\right)^6.
\end{align}
The integral will be non-zero for $m_1+m_2+m_3=0$.
The order of the $m$ value will determine the coefficient,
\beq
I_{3}^{\left(m_1,m_2,m_3\right)}=\sum_{m_1',m_2',m_3'}\left[D_{m_1m_1'}^{\left(2\right)}\left(\mathcal{R}\right)\right]^{*}\left[D_{m_2m_2'}^{\left(2\right)}\left(\mathcal{R}\right)\right]^{*}\left[D_{m_3m_3'}^{\left(2\right)}\left(\mathcal{R}\right)\right]^{*}\frac{\tilde{\zeta}_{m_1}\tilde{\zeta}_{m_2}\tilde{\zeta}_{m_3}}{\tilde{\zeta}_{m_1'}\tilde{\zeta}_{m_2'}\tilde{\zeta}_{m_3'}}K_3^{\left(m_1',m_2',m_3'\right)},
\eeq
where
\beq
K_{3}^{\left(m_{1}',m_{2}',m_{3}'\right)}=\begin{cases}
	J_{3}^{\left(2,-2,0\right)} & \left\{ m_{1}',m_{2}',m_{3}'\right\} =\left\{ 2,-2,0\right\} \\
	J_{3}^{\left(1,-1,0\right)} & \left\{ m_{1}',m_{2}',m_{3}'\right\} =\left\{ 1,-1,0\right\} \\
	J_{3}^{\left(1,1,-2\right)} & \left\{ m_{1}',m_{2}',m_{3}'\right\} =\left\{ 1,1,-2\right\}\ \textrm{or} \ \left\{ m_{1}',m_{2}',m_{3}'\right\} =\left\{ -1,-1,2\right\} \\
	J_{3}^{\left(0,0,0\right)} & \left\{ m_{1}',m_{2}',m_{3}'\right\} =\left\{ 0,0,0\right\} \\
	0 & \textrm{else}
\end{cases}
.\eeq

For the $S_zI_x$ interaction the decay is,
\begin{align}
&\left(\frac{\hbar\mu_0\gamma_e\gamma_N}{4\pi}\right)^3\tau^3 \left(I_3^{\left(1,1,1\right)}+I_3^{\left(1,1,-1\right)}+I_3^{\left(1,-1,1\right)}+I_3^{\left(-1,1,1\right)}+I_3^{\left(1,-1,-1\right)}+I_3^{\left(-1,1,-1\right)}+I_3^{\left(-1,-1,1\right)}+I_3^{\left(-1,-1,-1\right)}\right)\\\nonumber
&=-	\left(\frac{\hbar\mu_0\gamma_e\gamma_N}{4\pi}\right)^3\tau^3\frac{\pi  (257 \sin (2 \alpha )-18 \sin (6 \alpha ))}{4004 d^6}
\end{align}

\section{The power spectrum in our protocol}

In the last section of the main text we claim:
\beq
\left<\sum_{j=1}^N\tilde{G}_j^2\right>\propto\frac{nJ^2\tau}{2\pi}S\left(\delta\omega\right),
\eeq
where $S\left(\omega\right)$ is the power spectrum.
Here we would like to derive this relation explicitly.
We start with the average of $\tilde{G}_j^2$,
\begin{align}\label{PS1}
&\left<\sum_{j=1}^N\tilde{G}_j^2\right>=\sum_{j=1}^N\int\limits_0^\tau dt_1 \int\limits_0^\tau dt_2 h(t_1)h(t_2)\left<\sin\left(\omega_p t_1-\varphi_j(t_1)\right)\sin\left(\omega_p t_2-\varphi_j(t_2)\right) g^j_\pm\left(t_1\right) g^j_\pm\left(t_2\right)\right>\\\nonumber
=&-\frac{1}{4}\sum_{j=1}^N\int\limits_0^\tau dt_1 \int\limits_0^\tau dt_2 h(t_1)h(t_2)e^{i\omega_p(t_1+t_2)}\left<e^{-i\varphi_j(t_1)}g^j_\pm\left(t_1\right) e^{-i\varphi_j(t_2)}g^j_\pm\left(t_2\right)\right>\\\nonumber
&-\frac{1}{4}\sum_{j=1}^N\int\limits_0^\tau dt_1 \int\limits_0^\tau dt_2 h(t_1)h(t_2)e^{-i\omega_p(t_1+t_2)}\left<e^{i\varphi_j(t_1)}g^j_\pm\left(t_1\right)e^{i\varphi_j(t_2)} g^j_\pm\left(t_2\right)\right>
\\\nonumber
&+\frac{1}{4}\sum_{j=1}^N\int\limits_0^\tau dt_1 \int\limits_0^\tau dt_2 h(t_1)h(t_2)e^{i\omega_p(t_1-t_2)}\left<e^{-i\varphi_j(t_1)}g^j_\pm\left(t_1\right)e^{i\varphi_j(t_2)} g^j_\pm\left(t_2\right)\right>
\\\nonumber
&+\frac{1}{4}\sum_{j=1}^N\int\limits_0^\tau dt_1 \int\limits_0^\tau dt_2 h(t_1)h(t_2)e^{-i\omega_p(t_1-t_2)}\left<e^{i\varphi_j(t_1)}g^j_\pm\left(t_1\right)e^{-i\varphi_j(t_2)} g^j_\pm\left(t_2\right)\right>.
\end{align}
We recognize the terms inside the average as the stationary correlation functions:
\beq\label{PS2}
nJ^2C_2^{(m_1,m_2)}\left(t\right)\equiv nJ^2\tilde{\zeta}_{m_1}\tilde{\zeta}_{m_2}\int\frac{d^3r}{r^3}\int\frac{d^3r_0}{r_0^3}Y_2^{(m_1)}\left(\Omega\right)Y_2^{(m_2)}\left(\Omega_0\right)P(\bar{r},\bar{r}_0,t).
\eeq 
Eq. \eqref{PS1} can be rewritten in terms of \eqref{PS2} as 
\begin{align}\label{PS3}
&\frac{1}{nJ^2}\left<\sum_{j=1}^N\tilde{G}_j^2\right>=
-\frac{1}{4}\int\limits_0^\tau dt_1 \int\limits_0^\tau dt_2 h(t_1)h(t_2)e^{i\omega_p(t_1+t_2)}C_2^{(-1,-1)}\left(t_1-t_2\right)\\\nonumber
&-\frac{1}{4}\int\limits_0^\tau dt_1 \int\limits_0^\tau dt_2 h(t_1)h(t_2)e^{-i\omega_p(t_1+t_2)}C_2^{(1,1)}\left(t_1-t_2\right)
\\\nonumber
&+\frac{1}{4}\int\limits_0^\tau dt_1 \int\limits_0^\tau dt_2 h(t_1)h(t_2)e^{i\omega_p(t_1-t_2)}C_2^{(-1,1)}\left(t_1-t_2\right)
\\\nonumber
&+\frac{1}{4}\int\limits_0^\tau dt_1 \int\limits_0^\tau dt_2 h(t_1)h(t_2)e^{-i\omega_p(t_1-t_2)}C_2^{(1,-1)}\left(t_1-t_2\right).
\end{align}
If $\tau\ll \tau_D$ all these correlations in \eqref{PS3} are approximately constant in time and therefore
 \begin{align}\label{PS4}
 \frac{1}{nJ^2}\left<\sum_{j=1}^N\tilde{G}_j^2\right>\approx
 &-\frac{1}{4}I_2^{(-1,-1)}\int\limits_0^\tau dt_1 \int\limits_0^\tau dt_2 h(t_1)h(t_2)e^{i\omega_p(t_1+t_2)}
 -\frac{1}{4}I_2^{(1,1)}\int\limits_0^\tau dt_1 \int\limits_0^\tau dt_2 h(t_1)h(t_2)e^{-i\omega_p(t_1+t_2)}
 \\\nonumber
 &+\frac{1}{4}I_2^{(-1,1)}\int\limits_0^\tau dt_1 \int\limits_0^\tau dt_2 h(t_1)h(t_2)e^{i\omega_p(t_1-t_2)}
 +\frac{1}{4}I_2^{(1,-1)}\int\limits_0^\tau dt_1 \int\limits_0^\tau dt_2 h(t_1)h(t_2)e^{-i\omega_p(t_1-t_2)}\\\nonumber
 &=-\frac{\tau^2}{2}I_2^{(1,1)}\left(-\frac{2i}{\pi}\right)^2+\frac{\tau^2}{2}I_2^{(-1,1)}\left(-\frac{2i}{\pi}\right)\left(\frac{2i}{\pi}\right)=\frac{2\tau^2}{\pi^2}\frac{\pi  (13 \cos (4 \alpha )+51)}{1024 d^3}=\frac{\gamma_e^2 B_{rms}^2\tau^2}{nJ^2\pi^2}
 \end{align}
 If $\tau\gg\tau_D$ we can use:
\begin{align}
&\int\limits _{0}^{\tau}dt_1\int\limits _{0}^{\tau}dt_2 h(t_1)h(t_2)C(t_1-t_2 )=2\int\limits _{0}^{\tau}dt_1\int\limits _{-t_1/\sqrt{2}}^{t_1/\sqrt{2}}d\tilde{t}_2 h(t_1)h(t_1-\tilde{t}_2)C(\tilde{t}_2)
\label{Pulse1}\\
&\approx2\int\limits _{0}^{\tau}dt_{1}\int\limits _{-\infty}^{\infty}d\tilde{t}_{2}h\left(t_{1}\right)h\left(t_{1}-\tilde{t}_{2}\right)C\left(\tilde{t}_{2}\right)=2\int\frac{d\omega_{1}}{2\pi}\int\frac{d\omega_{2}}{2\pi}\int\frac{d\omega_{3}}{2\pi}\int\limits _{0}^{\tau}dt_{1}\int\limits _{-\infty}^{\infty}d\tilde{t}_{2}h_{\omega_{1}}e^{i\omega_{1}t_{1}}h_{\omega_{2}}e^{i\omega_{2}\left(t_{1}-\tilde{t}_{2}\right)}S_{\omega_{3}}e^{i\omega_{3}\tilde{t}_{2}}\label{Pulse2}
\end{align}
We shall solve all the integrals \eqref{PS3} at once:
\begin{align}
&\int\limits_0^\tau dt_1\int\limits_0^\tau dt_2h(t_1)h(t_2)e^{-im_1\omega_pt_1-im_2\omega_pt_2}C^{(m_1,m_2)}\left(t_1-t_2\right)\\\nonumber
&=2\int\frac{d\omega_{1}}{2\pi}\int\frac{d\omega_{2}}{2\pi}\int\frac{d\omega_{3}}{2\pi}\int\limits _{0}^{\tau}dt_{1}\int\limits _{-\infty}^{\infty}d\tilde{t}_{2}h_{\omega_{1}}e^{i\omega_{1}t_{1}}h_{\omega_{2}}e^{i\omega_{2}\left(t_{1}-\tilde{t}_{2}\right)}S^{(m_1,m_2)}_{\omega_{3}}e^{i\omega_{3}\tilde{t}_{2}}e^{-i(m_1+m_2)\omega_pt_1+im_2\omega_p\tilde{t}_2}
\\&=2\int\frac{d\omega_{1}}{2\pi}\int\frac{d\omega_{2}}{2\pi}\int d\omega_{3}\int\limits _{0}^{\tau}dt_{1}h_{\omega_{1}}e^{i\omega_{1}t_{1}}h_{\omega_{2}}e^{i\omega_{2}t_{1}}S^{(m_1,m_2)}_{\omega_{3}}e^{-i(m_1+m_2)\omega_pt_1}\delta\left(\omega_3-\omega_2+m_2\omega_p\right)\\
&=2\int\frac{d\omega_{1}}{2\pi}\int\frac{d\omega_{2}}{2\pi}\int\limits _{0}^{\tau}dt_{1}h_{\omega_{1}}h_{\omega_{2}}S^{(m_1,m_2)}_{\omega_2+m_2\omega_p}e^{i\left[\omega_1+\omega_2-(m_1+m_2)\omega_p\right]t_1}\\
&=2\int\frac{d\omega_{1}}{2\pi}\int\frac{d\omega_{2}}{2\pi}h_{\omega_{1}}h_{\omega_{2}}S^{(m_1,m_2)}_{\omega_2+m_2\omega_p}\frac{-i\left( e^{i\left[\omega_1+\omega_2-(m_1+m_2)\omega_p\right]\tau}-1\right)}{\omega_1+\omega_2-(m_1+m_2)\omega_p}\\
&=4\int\frac{d\omega_{1}}{2\pi}\int\frac{d\omega_{2}}{2\pi}h_{\omega_{1}}h_{\omega_{2}}S^{(m_1,m_2)}_{\omega_2+m_2\omega_p}e^{i\left[\omega_1+\omega_2-(m_1+m_2)\omega_p\right]\tau/2}\frac{ \sin\left[\left(\omega_1+\omega_2-(m_1+m_2)\omega_p\right)\tau\right]}{\omega_1+\omega_2-(m_1+m_2)\omega_p}\\
&\approx 4\pi\tau\int\frac{d\omega_{1}}{2\pi}\int\frac{d\omega_{2}}{2\pi}h_{\omega_{1}}h_{\omega_{2}}S^{(m_1,m_2)}_{\omega_2+m_2\omega_p}e^{i\left[\omega_1+\omega_2-(m_1+m_2)\omega_p\right]\tau/2}\delta\left(\omega_1+\omega_2-(m_1+m_2)\omega_p\right)\\
&=\frac{\tau}{\pi}\begin{cases}
\int d\omega_{1}\left|h_{\omega_{1}}\right|^2S^{(-m_2,m_2)}_{\omega_1-m_2\omega_p} & m_1=-m_2\\
\int d\omega_{1}\left|h_{\omega_{1}-m_2\omega_p}\right|^2S^{(m_2,m_2)}_{\omega_1-2m_2\omega_p} & m_1=m_2\\\label{PS5}
\end{cases}=\frac{\tau}{\pi}\int d\omega_{1}\left|h_{\omega_{1}}\right|^2S^{(\pm m_2,m_2)}_{\omega_1-m_2\omega_p}.
\end{align}
Substituting \eqref{PS5} into \eqref{PS3} we arrive at
\beq\label{PS6}
\left<\sum_{j=1}^N\tilde{G}_j^2\right>=\frac{\tau}{2\pi}\int d\omega_{1}\left|h_{\omega_{1}}\right|^2\left(S^{( -1,1)}_{\omega_1-\omega_p}-S^{( 1,1)}_{\omega_1-\omega_p}\right)
\eeq
Since $S\left(\omega-\omega_N\right)\propto S^{( -1,1)}_{\omega}-S^{( 1,1)}_{\omega}$, we can conclude
\beq\label{PS7}
\left<\sum_{j=1}^N\tilde{G}_j^2\right>=\frac{\tau}{2\pi}\int d\omega_{1}\left|h_{\omega_{1}}\right|^2S\left(\omega-\delta\omega-\omega_p\right)\approx\frac{\tau}{2\pi}S\left(\delta\omega\right),
\eeq
where we used the fact that $\left|h(\omega)\right|$ is a  narrow function that peaks at $\pm\omega_p$.
\bibliography{refs240919}